\documentclass[%
 reprint,
nofootinbib,
 amsmath,amssymb,
 aps,
]{revtex4-2}

\usepackage{graphicx}
\usepackage{multirow}
\usepackage{dcolumn}
\usepackage{bm}
\usepackage{pgf}
\usepackage{xcolor}
\usepackage{booktabs}
\usepackage{siunitx}
\definecolor{mypink}{RGB}{216, 18, 126}
\definecolor{myblue}{RGB}{0, 100, 162}
\usepackage[colorlinks=true, linkcolor=mypink, citecolor=myblue, urlcolor=myblue]{hyperref}
\usepackage{mathrsfs}
\numberwithin{equation}{section}
\renewcommand{\theequation}{\arabic{section}.\arabic{equation}}
\begin{document}

\preprint{APS/123-QED}

\title{A pre-merger-informed spectral-level ringdown inference framework for black-hole spectroscopy}

\author{Shitong Guo}
\email{shitongg@mail.nankai.edu.cn}%
\affiliation{School of Physics, Nankai University, 94 Weijin Road, Tianjin 300071, China}
\author{Yan-Gang Miao}%
\email{miaoyg@nankai.edu.cn (Corresponding author)}
\affiliation{School of Physics, Nankai University, 94 Weijin Road, Tianjin 300071, China}


\begin{abstract}

Black-hole spectroscopy aims to infer properties of the remnant spacetime from the quasinormal-mode (QNM) spectrum of the gravitational-wave ringdown signal.
In most implementations, however, this inference is performed with waveform models that already incorporate Kerr or other theory-specific QNM spectral relations, thereby entangling spectral measurement with remnant or beyond-Kerr parameter inference.
At the same time, conventional ringdown analyses commonly excise the pre-merger data, which in principle contain information about the excitation amplitudes and phases of the QNMs.
We introduce \texttt{SPRING} (\textit{Spectral-level Pre-merger-informed RINGdown inference}), a framework designed to separate ringdown spectral inference from theory-side interpretation while propagating pre-merger information through amplitude-scale estimation.
As a demonstration, we apply \texttt{SPRING} to GW250114 using an agnostic two-component damped-sinusoid (2DS) model for Kerr remnant inference.
\texttt{SPRING} improves the Bayesian support for the agnostic 2DS signal model relative to analyses that do not use pre-merger information, with an increase of \(\Delta\ln B\sim 5\)--\(10\).
The resulting remnant posterior remains closely consistent with the inspiral-merger-ringdown estimate, despite the extra freedom introduced by the second DS component.
This work bridges pre-merger information and ringdown inference, establishing a fully spectral-level route for future black-hole spectroscopy.
\end{abstract}

                              
\hypersetup{linkcolor=myblue}                             
\maketitle
\hypersetup{linkcolor=mypink}        


\section{Introduction}
\label{sec:introduction}

The gravitational-wave (GW) signal from a binary black-hole (BBH) coalescence is commonly divided into inspiral, merger, and ringdown stages.
The ringdown stage describes the post-merger relaxation of the newly formed remnant black hole (BH), and its signal carries information about the strongly curved spacetime near the remnant.
Once the remnant has settled sufficiently close to a stationary BH, the ringdown emission is expected to be well described by a superposition of exponentially damped sinusoidal modes, known as quasinormal modes (QNMs)
\cite{teukolskyPerturbationsRotatingBlack1973,
chandrasekharQuasinormalModesSchwarzschild1975,
leaverAnalyticRepresentationQuasinormal1985, 
kokkotasQuasiNormalModesStars1999,
bertiQuasinormalModesBlack2009}. 
The complex QNM frequencies encode the properties of the remnant spacetime and, in the Kerr case, are determined solely by the remnant mass and spin.\footnote{For astrophysical black holes, the electric charge is usually assumed to be negligible.}
This property of the QNM spectrum forms the basis of \textit{black-hole spectroscopy}
\cite{dreyerBlackHoleSpectroscopy2004,
baibhavMultimodeBlackHole2019,
bertiBlackHoleSpectroscopy2025,
bertiHowBlackHole2025,
carulloBlackHoleSpectroscopy2025},
whose primary goal is to measure the QNM spectrum encoded in the observed ringdown signal and use it to infer properties of the remnant spacetime or to test for possible deviations from general relativity (GR).

In practice, however, BH spectroscopy remains challenging.
The usable ringdown signal is short, its signal-to-noise ratio (SNR) is typically limited, and its spectral content is complicated by multimode structure
\cite{gieslerBlackHoleRingdown2019,
cotestaAnalysisRingdownOvertones2022,
capanoMultimodeQuasinormalSpectrum2023,
siegelGW231123RingdownInterpretation2025}
and nonlinear excitation effects
\cite{
cheungExtractingLinearNonlinear2024,
yangContributionNonlinearQuasinormal2025,
gieslerOvertonesNonlinearitiesBinary2025,
wangNonlinearVoiceGW2501142026}.
To extract more information from the available ringdown data and to stabilize the inference, a common strategy is to impose Kerr or other theory-specific spectral relations directly in the waveform model used for parameter inference
\cite{collaborationTestsGeneralRelativity2021,
collaborationTestsGeneralRelativity2021a,
siegelRingdownGW190521Hints2023,
gennariSearchingRingdownHigher2024,
cheungExtractingLinearNonlinear2024,
chungProbingQuadraticGravity2025,
collaborationBlackHoleSpectroscopy2025,
collaborationGW250114TestingHawkings2025}.
In such approaches, the spectral parameters appearing in the waveform model, namely the frequencies and damping times, are not inferred as independent data-side quantities.
Instead, they are tied from the outset to remnant or beyond-Kerr parameters through an assumed theoretical map.
This strategy is well motivated and has been central to current ringdown tests. 
Nevertheless, it introduces a structural limitation for BH spectroscopy. 
When this theoretical spectral relation is already imposed in the waveform model, the data-side measurement of the ringdown spectrum and the theory-side inference of remnant or beyond-Kerr parameters become entangled.
Consequently, the inferred remnant or beyond-Kerr properties reflect both the information extracted from the ringdown data and the assumptions built into the theoretical spectral map.
This makes it difficult to distinguish which features arise from the ringdown data and which are inherited from the imposed theoretical model.
It is therefore important to develop a framework in which spectral measurement and theoretical interpretation are explicitly separated.

A natural way to pursue this separation is to perform agnostic ringdown inference
\cite{collaborationTestsGeneralRelativity2016,
carulloObservationalBlackHole2019,
guoTheoryagnosticHierarchicalBayesian2025}. 
In this approach, the analysis is assumed to take place in the stationary relaxation regime, where the post-merger signal can be modeled as a superposition of damped-sinusoid (DS) components.
Each DS component is specified by a frequency, a damping time, an amplitude, and a phase, with the frequencies and damping times inferred directly from the ringdown data rather than constrained by a Kerr or other theory-specific relation.
This allows the ringdown spectral information to be measured before a theoretical interpretation is imposed.
However, this freedom comes at a price.
The dimensionality of the sampled parameter space grows rapidly with the number of DS components, which can affect sampler convergence and complicate the interpretation of the resulting posterior distribution
\cite{bertiBlackHoleSpectroscopy2025}.

Moreover, conventional ringdown analyses commonly excise the pre-merger data, even though this part of the signal can in principle inform the excitation amplitudes and phases of the QNMs
\cite{kamaretsosBlackholeHairLoss2012,
kamaretsosBlackHoleRingdownMemory2012}.
In such ringdown analyses, the amplitude--phase parameters are often assigned broad, uniform priors, leaving the inference to explore a weakly structured amplitude--phase space and potentially reducing the precision and stability of the spectral measurement.
This issue becomes especially important in agnostic ringdown inference, where the frequencies and damping times are also left free.

In this paper, we introduce \texttt{SPRING} (\textit{Spectral-level Pre-merger-informed RINGdown inference}), a framework designed to address these issues.
The central idea is to reorganize ringdown inference into three functionally distinct but interconnected layers.
For the GW event under consideration, we first choose a physically motivated QNM content to be tested.
The first layer then uses pre-merger information to estimate characteristic amplitude scales for these QNM components.
The second layer analyzes the ringdown data with an agnostic DS model containing the same number of DS components as the QNM components specified in the first layer.
The DS amplitude--phase parameters are assigned prior scales obtained from the first layer, guiding the ringdown inference toward a more physically informed region of parameter space, while the frequencies and damping times are inferred freely without imposing Kerr or other theory-specific spectral relations.
These amplitude--phase parameters are then analytically marginalized, yielding an explicit agnostic likelihood over the DS spectral parameters while reducing the effective dimensionality of the inference problem.
This agnostic DS spectral likelihood is then interpreted as the spectral likelihood function associated with the selected QNM content.
The third layer evaluates this spectral likelihood function on theory-predicted QNM spectra, thereby inferring remnant or beyond-Kerr parameters.
In this way, \texttt{SPRING} provides a modular spectral-level route for BH spectroscopy: pre-merger information enters only through the amplitude scale, while remnant or beyond-Kerr interpretation is deferred until after the spectral likelihood function has been constructed from the ringdown data.

We apply \texttt{SPRING} to GW250114
\cite{collaborationGW250114TestingHawkings2025,
collaborationBlackHoleSpectroscopy2025}
in a two-component damped-sinusoid (2DS) ringdown setting.
The results show that incorporating pre-merger amplitude information substantially improves the agnostic spectral inference.
In particular, the log Bayes factor increases by \(\Delta\ln B\sim 5\)--\(10\) relative to analyses with broad, uniform amplitude priors.
Using the resulting spectral likelihood function for Kerr remnant inference, we obtain a remnant posterior closely consistent with the inspiral-merger-ringdown (IMR) estimate, while the ringdown spectrum itself remains agnostically inferred.
We further test the dependence of the results on the pre-merger-informed amplitude scale and on the ringdown fitting start time, finding that both the spectral inference and the Kerr remnant inference remain stable under moderate variations of the amplitude scale and the fitting start time.

The paper is organized as follows.
In Sec.~\ref{sec:framework_overview}, we give an overview of the \texttt{SPRING} framework.
In Sec.~\ref{sec:premerger}, we describe the construction of the pre-merger-informed amplitude scale.
In Sec.~\ref{sec:spectral_inference}, we introduce a linear amplitude--phase representation of the ringdown DS template and derive the analytically marginalized spectral likelihood.
In Sec.~\ref{sec:theory_inference}, we present the theory-side inference formalism for inferring remnant or beyond-Kerr parameters from the spectral likelihood function.
In Sec.~\ref{sec:analysis_setup}, we specify the practical analysis setup used in this work.
In Sec.~\ref{sec:application_validation}, we apply the framework to GW250114, present the agnostic spectral-inference results, perform Kerr remnant inference, and carry out robustness tests.
We discuss the framework in Sec.~\ref{sec:discussion} and summarize our conclusions in Sec.~\ref{sec:conclusion}.

\section{Overview of the SPRING framework}
\label{sec:framework_overview}

\begin{figure*}
\centering
\includegraphics[
    width=0.92\textwidth,
    trim=10 33 10 20,
    clip
]{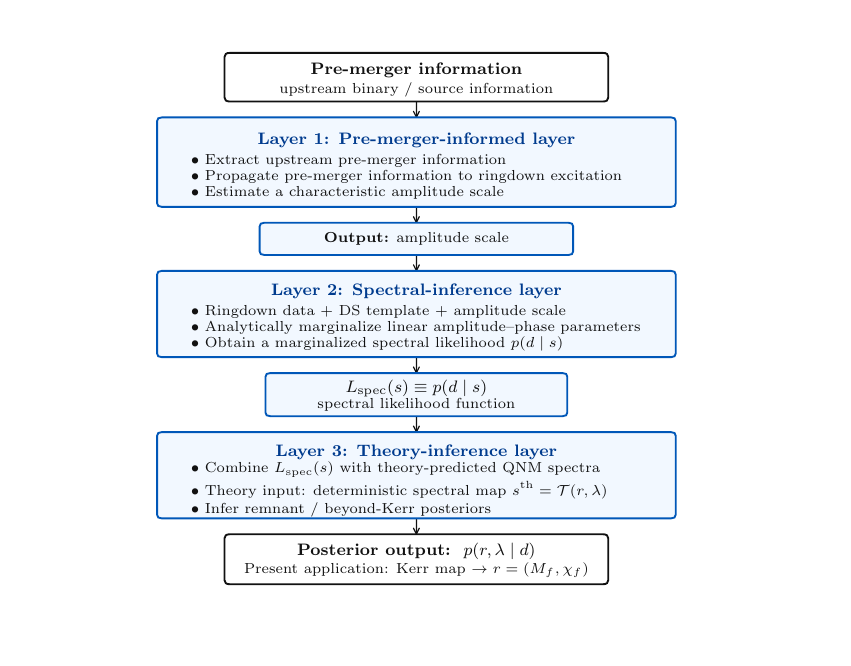}
\caption{
Schematic overview of the \texttt{SPRING} framework.
The information flow proceeds from the pre-merger information to an amplitude scale, then to the spectral likelihood function \(\mathcal{L}_{\rm spec}(s)\), and finally to remnant or beyond-Kerr inference through theory-predicted QNM spectra.
}
\label{fig:spring_framework}
\end{figure*}

The overall structure of \texttt{SPRING} is summarized in Fig.~\ref{fig:spring_framework}.
The framework is designed to make explicit the spectral-level object that connects ringdown data to remnant or beyond-Kerr parameters.
Rather than imposing a Kerr or theory-specific spectral relation directly in the waveform model, \texttt{SPRING} first constructs the spectral likelihood function through agnostic ringdown spectral inference.
Remnant or beyond-Kerr parameters are then inferred by evaluating this function at the spectra predicted by the corresponding theoretical map.

The framework consists of three layers.
The pre-merger-informed layer uses information carried by the pre-merger signal about the intrinsic binary parameters and extrinsic source parameters to define the amplitude scale, namely the prior scale assigned to the linear amplitude--phase coefficients of the ringdown model.
Here the intrinsic binary parameters include, for example, the masses and spins of the binary components, while the extrinsic source parameters include quantities relevant for the observed strain amplitude and phase, such as the luminosity distance, inclination, and reference phase.
This information is used only to estimate characteristic amplitude scales, not to impose deterministic amplitude values in the ringdown analysis.

The spectral-inference layer uses the ringdown data, the amplitude scale, and an agnostic DS template to construct the spectral likelihood function,
\begin{equation}
\mathcal{L}_{\rm spec}(s)
\equiv
p(d\mid s),
\end{equation}
where the dependence on the adopted amplitude scale is kept implicit.
Here \(p(d\mid s)\) denotes the marginalized spectral likelihood obtained after analytically marginalizing over the linear amplitude--phase coefficients.
The variable \(d\) denotes the ringdown data used for spectral inference, and \(s\) denotes the frequencies and damping times of the DS components.

The theory-inference layer combines \(\mathcal{L}_{\rm spec}(s)\) with a theory-side spectral prediction \(p(s\mid r,\lambda)\), where \(r\) denotes remnant parameters and \(\lambda\) denotes possible beyond-Kerr parameters.
In the deterministic case, this prediction reduces to a spectral map \(s^{\rm th}=\mathcal{T}(r,\lambda)\), and the posterior is obtained by evaluating the spectral likelihood function at the theory-predicted spectrum, \(\mathcal{L}_{\rm spec}(s^{\rm th})\), together with the prior on \((r,\lambda)\), as detailed in Sec.~\ref{sec:theory_inference}.

The information flow can therefore be written schematically as
\begin{equation}
\begin{aligned}
&\text{pre-merger information}
\longrightarrow
\text{amplitude scale} \\
&\longrightarrow
\mathcal{L}_{\rm spec}(s)
\longrightarrow
p(r,\lambda\mid d).
\end{aligned}
\end{equation}
In the present application, we use the Kerr QNM spectral map
\cite{Stein:2019mop}
and infer the remnant parameters \(r=(M_f,\chi_f)\), where \(M_f\) is the detector-frame remnant BH mass and \(\chi_f\) is the dimensionless remnant spin.
The formalism itself, however, applies to any theory that predicts QNM spectra as functions of remnant and, when present, beyond-Kerr parameters.

\section{Pre-merger-informed amplitude scale}
\label{sec:premerger}

We now turn to the concrete methodology of \texttt{SPRING}, beginning with the pre-merger-informed layer.
The role of this layer is to propagate pre-merger information into a characteristic amplitude scale for the linear amplitude--phase coefficients used in the spectral-inference layer.

We first specify the source of the pre-merger information used to construct the amplitude scale.
The required pre-merger information may be obtained from either an inspiral-only posterior or an IMR posterior.
In either case, the pre-merger information is used only to define a characteristic amplitude scale, not to impose deterministic amplitude values in the ringdown analysis.
The subsequent spectral inference and theory-side inference are therefore not expected to be dominated by the choice of pre-merger information source.
This expectation can be assessed by comparing amplitude scales constructed from different pre-merger inputs; in the application below, we compare IMR posterior samples with inspiral-only posterior samples.
The robustness of the analysis can also be examined by applying moderate variations to the adopted amplitude scale.

For each pre-merger posterior sample, we use \texttt{pyseobnr}
\cite{mihaylovPySEOBNRSoftwarePackage2023},
a Python package for GW modeling within the effective-one-body (EOB) formalism
\cite{buonannoEffectiveOnebodyApproach1999,
buonannoTransitionInspiralPlunge2000,
damourCoalescenceTwoSpinning2001,
buonannoTransitionInspiralPlunge2006,
buonannoSourcesGravitationalWaves2015,
Ramos-Buades:2023ehm,
Pompili:2023tna,
ng8w-98sz},
to generate a set of time-domain waveform modes,
\begin{equation}
\{h_{\ell m}(t)\}.
\end{equation}
Here the subscripts \(\ell\) and \(m\) label the angular waveform modes \((\ell,m)\); the overtone index \(n\) is introduced below when specifying the QNM basis.

Before extracting the amplitude scale, we first construct a target waveform to be fitted by the QNM basis.
The target waveform is built from a selected set of waveform modes denoted by \(\mathcal{M}_{\rm tar}\).
For nearly equal-mass, nonprecessing, quasicircular binaries such as GW150914
\cite{abbottPropertiesBinaryBlack2016,
abbottObservationGravitationalWaves2016}
or GW250114, the \(\ell=|m|=2\) modes are expected to dominate the ringdown signal.
In the present application, we focus on the prograde \((\ell,m)=(2,2)\) waveform mode and neglect the corresponding retrograde contribution, which is generally less strongly excited and subdominant for such binary systems
\cite{fortezaSpectroscopyBinaryBlack2020,
Dhani:2020nik,
Li_2022}.
We further support this choice for the event considered here using the pre-merger-informed waveform-mode contribution diagnostics presented in Appendix~\ref{app:dominant_mode}.
Accordingly, in this work we take
\begin{equation}
\mathcal{M}_{\rm tar}=\{(2,2)\}.
\end{equation}

The selected waveform modes are then combined with spin-weighted spherical harmonics at the source orientation to construct the projected complex strain used as the target waveform for QNM amplitude scale extraction.
Given \(\mathcal{M}_{\rm tar}\), we construct the projected complex target waveform as
\begin{equation}
h_{\rm tar}(t)
=
\sum_{(\ell,m)\in \mathcal{M}_{\rm tar}}
{}_{-2}Y_{\ell m}
\left(
\iota,
\frac{\pi}{2}-\phi_{\rm ref}
\right)
h_{\ell m}(t).
\label{eq:htar_projection}
\end{equation}
Here \(\iota\) is the inclination, and \(\phi_{\rm ref}\) is the reference phase passed to \texttt{pyseobnr}.\footnote{
Different waveform conventions may differ by phase or azimuthal-angle definitions.
In the present implementation, the posterior value of \texttt{phase} is passed directly as the \texttt{pyseobnr} input parameter \(\phi_{\rm ref}\), and the spin-weighted spherical harmonics are evaluated at azimuth \(\pi/2-\phi_{\rm ref}\), following the \texttt{pyseobnr}/\texttt{LAL} convention.
}

For each pre-merger posterior sample, the generated waveform modes also define a sample-dependent peak time.
We compute this peak time as
\begin{equation}
t_{\rm peak}
=
\arg\max_t
\sum_{(\ell,m)\in \mathcal{M}_{\rm peak}}
|h_{\ell m}(t)|^2 ,
\label{eq:tpeak_mode_norm}
\end{equation}
where \(\mathcal{M}_{\rm peak}\) is the mode set used for the peak definition.
This definition is independent of the observer's orientation and is commonly used as a reference time in numerical-relativity (NR) studies
\cite{mitmanProbingRingdownPerturbation2025,
Gao_2025}.
The ringdown fitting start time is then chosen as
\begin{equation}
t_0
=
t_{\rm peak}
+
\kappa_0\,t_{M_f} ,
\label{eq:t0_premerger}
\end{equation}
with
\begin{equation}
t_{M_f}
=
\frac{G M_f}{c^3},
\label{eq:tMf_def}
\end{equation}
where \(\kappa_0\) is a dimensionless offset.
Here \(M_f\) is the detector-frame remnant mass associated with the same pre-merger posterior sample used to generate the waveform modes, so that the remnant-mass time unit \(t_{M_f}\) in Eq.~\eqref{eq:tMf_def} is consistently defined for each sample.
Thus, in the pre-merger-informed layer, both \(t_{\rm peak}\) and \(t_{M_f}\) are evaluated sample by sample, so the fitting start time \(t_0\) is also sample-dependent.

Having constructed the target waveform and chosen the ringdown fitting start time, we next specify the physically motivated set of QNM components used for the amplitude fit.
We denote this set by \(\mathcal{J}\).
The choice of \(\mathcal{J}\) is guided by NR studies or by existing data-analysis results for the relevant ringdown signal.
Starting from \(t_0\), the projected target waveform is fitted with the selected QNM basis,
\begin{equation}
h_{\rm tar}(t)
\simeq
\sum_{j\in\mathcal{J}}
\mathcal{A}_j
\exp\left[-i\omega_j(t-t_0)\right],
\qquad
t\geq t_0 ,
\label{eq:htar_qnm_fit}
\end{equation}
where \(j=(\ell,m,n)\) labels a QNM component, \(\mathcal{A}_j\) is the complex amplitude of component \(j\), and \(\omega_j\) is the corresponding complex QNM frequency.
The frequencies \(\omega_j\) are evaluated at the remnant mass and spin associated with each pre-merger posterior sample, and are used only within the pre-merger-informed layer to construct the QNM basis for the amplitude fit.

The complex amplitudes are obtained by ordinary complex least squares.
Over the fitting time interval, we evaluate the target waveform at discrete times \(t_a\) and arrange the values \(h_{\rm tar}(t_a)\) into a vector \(\mathbf{h}_{\rm tar}\).
The complex amplitudes of the selected QNM components are collected into the coefficient vector \(\boldsymbol{\mathcal{A}}\).
The corresponding QNM fitting matrix \(\mathbf{B}\) is obtained by evaluating the selected QNM basis functions at the same discrete times,
\begin{equation}
B_{aj}
=
\exp\left[-i\omega_j(t_a-t_0)\right],
\label{eq:layer1_design_matrix}
\end{equation}
where \(a\) labels the time sample.
The least-squares fit is therefore performed for the linear system
\(\mathbf{h}_{\rm tar}\simeq \mathbf{B}\boldsymbol{\mathcal{A}}\).
We compute
\begin{equation}
\hat{\boldsymbol{\mathcal{A}}}
=
\arg\min_{\boldsymbol{\mathcal{A}}}
\sum_a
\left|
(\mathbf{h}_{\rm tar})_a
-
\sum_{j\in\mathcal{J}} B_{aj}\mathcal{A}_j
\right|^2 .
\label{eq:complex_lstsq_layer1}
\end{equation}
Repeating this procedure over the target waveforms generated from the pre-merger posterior samples produces a distribution of fitted complex amplitudes.

In the present application, we use a two-component QNM fit to the
\((\ell,m)=(2,2)\) target waveform, with
\(\mathcal{J}=\{(2,2,0),(2,2,1)\}\), referred to compactly as the \(220\) and
\(221\) components.
The fit therefore gives the complex amplitudes \(\mathcal{A}_{220}\) and \(\mathcal{A}_{221}\).
This choice is motivated by the expected ringdown hierarchy for a GW250114-like system: the \(220\) component is the longest-lived fundamental mode and dominates at late times, while NR fits to similar systems identify the \(221\) overtone as the next strongest mode
\cite{buonannoInspiralMergerRingdown2007,
londonModelingRingdownFundamental2014,
gieslerBlackHoleRingdown2019,
cheungExtractingLinearNonlinear2024,
pacilioFlexibleMappingRingdown2024,
mitmanProbingRingdownPerturbation2025,
zertucheHighPrecisionRingdownSurrogate2025}.

The output of the pre-merger-informed layer is therefore an empirically estimated amplitude scale for each selected QNM component. 
We define this scale as the median of the fitted complex-amplitude magnitudes \(|\mathcal{A}_j|\) over the ensemble of target waveforms generated from the pre-merger posterior samples.
In the spectral-inference layer, this amplitude scale is used as the prior scale for the linear amplitude--phase coefficients of the corresponding agnostic DS components.

\section{Analytically marginalized spectral inference}
\label{sec:spectral_inference}

In this section, we construct the agnostic spectral-inference layer of \texttt{SPRING}.
In Sec.~\ref{sec:ds_linear_model}, we rewrite the ringdown DS template in terms of linear amplitude--phase coefficients, making explicit that the waveform is linear in these coefficients for fixed frequencies and damping times.
In Sec.~\ref{sec:analytic_marginalization}, we analytically marginalize over the linear amplitude--phase coefficients, with the amplitude scales from the pre-merger-informed layer entering naturally as prior scales.

\subsection{Damped-sinusoid model and linear representation}
\label{sec:ds_linear_model}

We consider an agnostic DS representation of the ringdown signal in a time interval \(t\in[t_0,t_0+T]\), where the remnant is expected to have entered the stationary relaxation regime.
Within this interval, the post-merger strain is modeled using a DS template:
\begin{equation}
\begin{aligned}
\mathcal{H}(t)
&= h_+(t)-i h_\times(t)  \\
&=
\sum_{k=1}^{K}
A_k\,
\Theta(t-t_0)\,
e^{-(t-t_0)/\tau_k}
e^{i[-2\pi f_k(t-t_0)+\phi_k]} .
\end{aligned}
\label{eq:H_DS_complex}
\end{equation}
Here the step function \(\Theta(t-t_0)\) restricts the template to the ringdown interval \(t\ge t_0\).
For the \(k\)th agnostic DS component, \(f_k>0\) and \(\tau_k>0\) denote the frequency and damping time, while \(A_k\ge0\) and \(\phi_k\) denote the amplitude and phase.
We collect the ordered spectral variables as
\begin{equation}
s \equiv (f_1,\tau_1,\ldots,f_K,\tau_K).
\label{eq:spectral_variables_def}
\end{equation}
The index \(k=1,\ldots,K\) labels DS components and should be distinguished from the QNM component label \(j=(\ell,m,n)\) used in the pre-merger-informed layer.
In \texttt{SPRING}, \(K\) is chosen to match the number of QNM components selected in the pre-merger-informed layer, but the DS spectral variables \((f_k,\tau_k)\) are inferred agnostically.

For each DS component, the amplitude and phase in Eq.~\eqref{eq:H_DS_complex} are rewritten in terms of real linear amplitude--phase coefficients,
\begin{equation}
A_k e^{i\phi_k}=p_k+i q_k .
\label{eq:pq_def}
\end{equation}
Defining
\begin{equation}
\begin{aligned}
b_{\cos,k}(t)
&\equiv
\Theta(t-t_0)e^{-(t-t_0)/\tau_k}
\cos[2\pi f_k(t-t_0)], \\
b_{\sin,k}(t)
&\equiv
\Theta(t-t_0)e^{-(t-t_0)/\tau_k}
\sin[2\pi f_k(t-t_0)] ,
\end{aligned}
\label{eq:basis_defs}
\end{equation}
the two polarizations can be written as
\begin{align}
h_+(t)
&=
\sum_{k=1}^{K}
\left[
p_k b_{\cos,k}(t)+q_k b_{\sin,k}(t)
\right],
\label{eq:hplus_pq}
\\
h_\times(t)
&=
-\sum_{k=1}^{K}
\left[
q_k b_{\cos,k}(t)-p_k b_{\sin,k}(t)
\right].
\label{eq:hcross_pq}
\end{align}
For fixed \(s\), the waveform template is therefore linear in the real coefficients \(p_k\) and \(q_k\).

The two polarizations in Eqs.~\eqref{eq:hplus_pq}--\eqref{eq:hcross_pq}
define the polarization waveform template.
For each detector \(X\), this template is projected through the detector antenna
response to obtain the detector-frame model strain,
\begin{equation}
h^{(X)}(t)
=
F_{+}^{(X)} h_{+}(t)
+
F_{\times}^{(X)} h_{\times}(t),
\label{eq:detector_strain_def}
\end{equation}
where \(F_{+}^{(X)}\) and \(F_{\times}^{(X)}\) are the antenna-pattern functions for detector \(X\), determined by the source sky location and polarization angle.
Substituting Eqs.~\eqref{eq:hplus_pq} and \eqref{eq:hcross_pq} into Eq.~\eqref{eq:detector_strain_def},
the detector-frame model strain can be written explicitly as
\begin{equation}
\begin{aligned}
h^{(X)}(t)
&=
\sum_{k=1}^{K}
\Bigl[
\bigl(
F_{+}^{(X)} b_{\cos,k}(t)
+
F_{\times}^{(X)} b_{\sin,k}(t)
\bigr)p_k \\
&\qquad+
\bigl(
F_{+}^{(X)} b_{\sin,k}(t)
-
F_{\times}^{(X)} b_{\cos,k}(t)
\bigr)q_k
\Bigr].
\end{aligned}
\label{eq:detector_strain_linear}
\end{equation}
Thus, for each detector and each choice of spectral variables, the detector-frame model strain remains linear in the amplitude--phase coefficients.

After projecting the polarization waveform template onto each detector, we evaluate Eq.~\eqref{eq:detector_strain_linear} at the same time samples as the detector data within the analysis time interval.
Concatenating the model strain from all detectors gives the stacked detector-frame waveform template,
\begin{equation}
h = M(s)\alpha .
\label{eq:stacked_template}
\end{equation}
The stacked detector data are then modeled as the sum of this template and detector noise,
\begin{equation}
d = h+n = M(s)\alpha+n,
\label{eq:linear_model}
\end{equation}
where
\begin{equation}
\alpha\equiv(p_1,q_1,\ldots,p_K,q_K)^T .
\label{eq:alpha_def}
\end{equation}
Here \(d\) denotes the stacked detector data vector, obtained by concatenating the time-domain strain samples from all detectors in the analysis time interval; 
\(h\) is the corresponding stacked detector-frame waveform template;
\(n\) is the corresponding stacked noise vector; 
and \(M(s)\) is the DS design matrix constructed from the DS basis functions and the detector responses.

\subsection{Analytic marginalization}
\label{sec:analytic_marginalization}

We assume Gaussian detector noise with covariance \(C\). 
The corresponding Gaussian likelihood is
\begin{equation}
p(d\mid \alpha,s)
\propto
\exp\left[
-\frac12(d-M(s)\alpha)^T C^{-1}(d-M(s)\alpha)
\right].
\label{eq:likelihood_alpha_s}
\end{equation}
The amplitude scales obtained in the pre-merger-informed layer enter through a zero-mean Gaussian prior on the linear coefficients,
\begin{equation}
\alpha\sim\mathcal{N}(0,\Lambda),
\qquad
\Lambda=
\mathrm{diag}
\left(
\sigma_1^2,\sigma_1^2,\ldots,\sigma_K^2,\sigma_K^2
\right).
\label{eq:Lambda_diag}
\end{equation}
Here \(\sigma_k\) is the amplitude scale assigned to the \(k\)th DS component, so that the two linear coefficients \((p_k,q_k)\) have the same prior variance \(\sigma_k^2\).

In the present application, the selected QNM content is \(\{220,221\}\), with the \(220\) fundamental mode longer lived than the \(221\) overtone.
We therefore order the two DS components by decreasing damping time and assign the \(220\) scale to the longer-lived component and the \(221\) scale to the shorter-lived component.
With the median definition of the amplitude scale used in Sec.~\ref{sec:premerger}, this gives
\begin{equation}
\sigma_1=\mathrm{median}\left(|\mathcal{A}_{220}|\right),
\qquad
\sigma_2=\mathrm{median}\left(|\mathcal{A}_{221}|\right),
\end{equation}
where \(k=1\) denotes the longer-lived DS component and \(k=2\) denotes the shorter-lived DS component.
This convention provides one amplitude scale for each agnostic DS component.

This prior is isotropic in each \((p_k,q_k)\) plane.
In polar amplitude--phase variables, it induces
\begin{equation}
\pi(A_k,\phi_k)
=
\frac{A_k}{2\pi\sigma_k^2}
\exp\left[
-\frac{A_k^2}{2\sigma_k^2}
\right],
\quad
A_k\ge0,\quad
\phi_k\in[0,2\pi).
\label{eq:prior_A_phi}
\end{equation}
Thus, for each DS component, the phase is uniform and the amplitude follows a Rayleigh distribution with scale \(\sigma_k\).
The amplitude prior therefore peaks at the pre-merger-informed scale \(\sigma_k\), providing a soft preference for amplitudes near this scale and guiding the ringdown inference toward a more physically informed region of parameter space.
This differs from treating the amplitudes as free in the sense of assigning broad, uniform amplitude priors, for which the amplitude is largely uninformed.
At the same time, the amplitude scale is not a deterministic constraint: the prior retains support over all \(A_k\ge0\), and the amplitude is not fixed to a prescribed value.

The marginalized spectral likelihood for the DS model is obtained by integrating over the linear amplitude--phase coefficients,
\begin{equation}
p(d\mid s)
=
\int
d\alpha\;
p(d\mid \alpha,s)\,
p(\alpha).
\label{eq:marg_like_def}
\end{equation}
Here \(p(\alpha)\) denotes the Gaussian prior specified above; we assume \(p(\alpha\mid s)=p(\alpha)\) once the amplitude scales are fixed.
Because both \(p(d\mid \alpha,s)\) and \(p(\alpha)\) are Gaussian in the linear coefficients \(\alpha\), the integral in Eq.~\eqref{eq:marg_like_def} can be evaluated analytically
\cite{Prix2016Ringdown,
hoggDataAnalysisRecipes2020}.
Introducing the whitened data $d_w$ and whitened DS design matrix $V$,
\begin{equation}
d_w=L^{-1}d,\qquad V=L^{-1}M(s),\qquad C=LL^T,
\end{equation}
and defining
\begin{equation}
\chi_0^2=d_w^T d_w,\quad
\mu=V^T d_w,\quad
S_0=V^T V,\quad
S=S_0+\Lambda^{-1},
\end{equation}
one obtains
\begin{equation}
\begin{aligned}
\log p(d\mid s)
=
&-\frac{1}{2}\chi_0^2
+
\frac{1}{2}\mu^T S^{-1}\mu
-
\frac{1}{2}\log\det S  \\
&-
\frac{1}{2}\log\det\Lambda
-
\frac{1}{2}\log\det C
-
\frac{N}{2}\log(2\pi).
\end{aligned}
\label{eq:log_marg_like}
\end{equation}
Here \(N\) is the dimension of the stacked data vector \(d\), i.e., the total number of time-domain strain samples obtained by concatenating all detectors over the analysis interval \(t\in[t_0,t_0+T]\).

We define the \textit{spectral likelihood function} as
\begin{equation}
\mathcal{L}_{\rm spec}(s)\equiv p(d\mid s),
\label{eq:Lspec_def}
\end{equation}
with the dependence on the adopted amplitude scales kept implicit.
The function \(\mathcal{L}_{\rm spec}(s)\) is the output of the spectral-inference layer: an explicitly evaluable likelihood function over the ringdown spectral variables obtained from the analytic marginalization above.

In \texttt{SPRING}, this spectral likelihood function provides the interface between the data-side agnostic spectral inference and its theory-side interpretation.
For a chosen QNM content, the agnostic DS model is used to infer the same number of spectral components.
The correspondence between DS components and the selected QNM components is fixed by the damping-time ordering convention and by the pre-merger-informed amplitude-scale assignment.
Under this identification, the marginalized likelihood \(p(d\mid s)\) is interpreted as the spectral likelihood function, \(\mathcal{L}_{\rm spec}(s)\equiv p(d\mid s)\), associated with the selected QNM content.
In the theory-inference layer, this likelihood function is then evaluated on theory-predicted QNM spectra.

\section{Theory-side inference from the spectral likelihood}
\label{sec:theory_inference}

We now formulate the theory-inference layer built on top of the spectral likelihood function constructed in Sec.~\ref{sec:spectral_inference}.
This layer infers remnant or beyond-Kerr parameters by evaluating \(\mathcal{L}_{\rm spec}(s)\) on spectra predicted by a given theoretical model.

For each event \(e\), we denote the observed data by \(d_e\), the remnant parameters by \(r_e\), the spectral variables by \(s_e\), and the linear amplitude--phase coefficients by \(\alpha_e\). 
The beyond-Kerr parameters are denoted by \(\lambda\), which may be shared across events depending on the theory under consideration.
In the theory-inference layer, the variables \(s_e\) are interpreted as the physical spectral quantities of the selected QNM content, rather than merely as the unlabeled agnostic DS coordinates used in the spectral-inference layer.
Once the component-identification convention is specified, the agnostic DS component labels are associated with the corresponding QNM labels.
The coefficients \(\alpha_e\) describe the associated QNM excitation information from the preceding binary dynamics.

Our starting point is a conditional-independence assumption appropriate to the ideal linear ringdown regime:
\begin{equation}
d_e \perp (r_e,\lambda)\mid (\alpha_e,s_e),
\label{eq:cond_indep_data}
\end{equation}
or equivalently,
\begin{equation}
p(d_e\mid \alpha_e,s_e,r_e,\lambda)=p(d_e\mid \alpha_e,s_e).
\label{eq:cond_indep_data_likelihood}
\end{equation}
This means that, once the ringdown waveform representation \((\alpha_e,s_e)\) is specified, the data contain no additional dependence on the remnant or beyond-Kerr parameters.

With these definitions in place, we now develop the formalism that connects the spectral likelihood function to posterior inference on remnant or beyond-Kerr parameters.
Starting from Bayes' theorem, the target posterior for a single event is
\begin{equation}
p(r_e,\lambda\mid d_e)
\propto
p(d_e\mid r_e,\lambda)\,
p(r_e,\lambda).
\label{eq:theory_posterior_start}
\end{equation}
The factor \(p(d_e\mid r_e,\lambda)\) is the event likelihood for the theory-side parameters.
It can be expanded by inserting the ringdown variables \((\alpha_e,s_e)\),
\begin{equation}
p(d_e\mid r_e,\lambda)
=
\int ds_e\,d\alpha_e\;
p(d_e\mid \alpha_e,s_e,r_e,\lambda)\,
p(\alpha_e,s_e\mid r_e,\lambda).
\label{eq:event_likelihood_expand}
\end{equation}
Using Eq.~\eqref{eq:cond_indep_data_likelihood} and factorizing
\(p(\alpha_e,s_e\mid r_e,\lambda)\), this becomes
\begin{equation}
\begin{aligned}
p(d_e\mid r_e,\lambda)
&=
\int ds_e\,d\alpha_e\;
p(d_e\mid \alpha_e,s_e)  \\
&\quad\times
p(\alpha_e\mid s_e,r_e,\lambda)\,
p(s_e\mid r_e,\lambda).
\end{aligned}
\label{eq:event_likelihood_general}
\end{equation}

We now specialize to the deterministic spectral maps used in this work.
For such a theory-side prediction, the remnant parameters, together with any beyond-Kerr parameters when present, determine a unique QNM spectrum,
\begin{equation}
s_e^{\rm th}
=
\mathcal{T}(r_e,\lambda),
\label{eq:theory_spectrum_map}
\end{equation}
where \(\mathcal{T}\) denotes the theory-side spectral map.
Equivalently,
\begin{equation}
p(s_e\mid r_e,\lambda)
=
\delta\!\left(s_e-s_e^{\rm th}\right).
\label{eq:deterministic_spectral_map}
\end{equation}
Equation~\eqref{eq:event_likelihood_general} then reduces to
\begin{equation}
p(d_e\mid r_e,\lambda)
=
\int d\alpha_e\;
p(d_e\mid \alpha_e,s_e^{\rm th})\,
p(\alpha_e\mid s_e^{\rm th},r_e,\lambda).
\label{eq:event_likelihood_deterministic_alpha}
\end{equation}

For the deterministic spectral map above, the only conditional distribution of \(\alpha_e\) required in the event likelihood is \(p(\alpha_e\mid s_e^{\rm th},r_e,\lambda)\), evaluated at the theory-predicted spectral point \(s_e^{\rm th}=\mathcal{T}(r_e,\lambda)\).
In the parameter region considered, we assume that the selected QNM spectrum provides a sufficient spectral representation of the theory-side parameters used in the inference.
Once \(s_e^{\rm th}\) has been specified, the additional conditioning on \((r_e,\lambda)\) is therefore redundant, and we write
\begin{equation}
p(\alpha_e\mid s_e^{\rm th},r_e,\lambda)
=
p(\alpha_e\mid s_e^{\rm th}).
\label{eq:amp_spectral_closure_deterministic}
\end{equation}
Thus the excitation coefficients \(\alpha_e\) are conditioned on the theory-predicted QNM spectrum itself, rather than additionally on the parameters \((r_e,\lambda)\) that generate this spectral point.

With this prescription, Eq.~\eqref{eq:event_likelihood_deterministic_alpha} becomes
\begin{equation}
p(d_e\mid r_e,\lambda)
=
\int d\alpha_e\;
p(d_e\mid \alpha_e,s_e^{\rm th})\,
p(\alpha_e\mid s_e^{\rm th}).
\label{eq:event_likelihood_deterministic_margalpha}
\end{equation}
The integral over \(\alpha_e\) is precisely the spectral likelihood function evaluated at the theory-predicted spectrum,
\begin{equation}
\mathcal{L}_{{\rm spec},e}(s_e^{\rm th})
\equiv
p(d_e\mid s_e^{\rm th})
=
\int d\alpha_e\;
p(d_e\mid \alpha_e,s_e^{\rm th})\,
p(\alpha_e\mid s_e^{\rm th}).
\label{eq:event_lspec_deterministic}
\end{equation}
Therefore,
\begin{equation}
p(d_e\mid r_e,\lambda)
=
\mathcal{L}_{{\rm spec},e}\!\left(s_e^{\rm th}\right).
\label{eq:event_likelihood_deterministic}
\end{equation}
Substituting this into Eq.~\eqref{eq:theory_posterior_start}, we obtain
\begin{equation}
p(r_e,\lambda\mid d_e)
\propto
p(r_e,\lambda)\,
\mathcal{L}_{{\rm spec},e}\!\left(s_e^{\rm th}\right).
\label{eq:theory_posterior_deterministic}
\end{equation}

The construction extends directly to multiple events.
Assuming that the detector data from different events are conditionally independent once the parameters of each event are specified, the joint posterior for a shared beyond-Kerr parameter \(\lambda\) and event-dependent remnant parameters \(\{r_e\}\) is
\begin{equation}
p(\{r_e\},\lambda\mid \{d_e\})
\propto
p(\{r_e\},\lambda)
\prod_e
p(d_e\mid r_e,\lambda).
\label{eq:multi_event_posterior_start}
\end{equation}
For deterministic spectral maps, substituting Eq.~\eqref{eq:event_likelihood_deterministic} into Eq.~\eqref{eq:multi_event_posterior_start} gives
\begin{equation}
\begin{aligned}
p(\{r_e\},\lambda\mid \{d_e\})
&\propto
p(\{r_e\},\lambda)
\prod_e
\mathcal{L}_{{\rm spec},e}(s_e^{\rm th}).
\end{aligned}
\label{eq:multi_event_spectral}
\end{equation}

In this way, the theory-inference layer completes the spectral-level route from ringdown data to remnant or beyond-Kerr parameters: the data enter through the spectral likelihood function, and the theory enters through its prediction for the QNM spectrum.

\section{Analysis setup}
\label{sec:analysis_setup}

This section specifies the practical analysis setup used in the GW250114 application.
In Sec.~\ref{sec:data_preprocessing}, we describe the data preprocessing procedure.
In Sec.~\ref{sec:inference_setup}, we specify the inference settings, including the priors, sampler settings, and the Kerr QNM calculation used in the theory-side inference.

\subsection{Data preprocessing}
\label{sec:data_preprocessing}

For the GW250114 analysis, we use a \(64\,{\rm s}\) strain segment centered on the event GPS time.
We start from the publicly available \(16384\,{\rm Hz}\) strain data and downsample the time series to \(4096\,{\rm Hz}\).
The resulting Nyquist frequency is therefore \(2048\,{\rm Hz}\).
We then apply a bandpass filter over the frequency range \(20\)--\(2043\,{\rm Hz}\), retaining the frequency band relevant for the ringdown analysis while avoiding the Nyquist edge.

The noise power spectral density (PSD) is estimated from the same \(64\,{\rm s}\) data segment after downsampling and bandpass filtering.
Following the conditioning strategy used in previous time-domain ringdown analyses
\cite{isiAnalyzingBlackholeRingdowns2021,
siegelAnalyzingBlackholeRingdowns2025}, we then apply a PSD-censoring procedure: the PSD is inflated outside the effective frequency range \(20\)--\(1830\,{\rm Hz}\) used in the likelihood.
This strongly downweights the contribution of frequencies outside the trusted band through the noise covariance.

The time-domain covariance matrix \(C\) entering the Gaussian likelihood in Eq.~\eqref{eq:likelihood_alpha_s} is constructed from this conditioned PSD.

\subsection{Inference setup}
\label{sec:inference_setup}

Throughout this work, we perform a joint analysis of the H1 and L1 strain data.
The detector antenna-pattern functions are fixed using the sky location and polarization angle reported for GW250114
\cite{collaborationGW250114TestingHawkings2025}:
right ascension \(2.333\,{\rm rad}\), declination \(0.190\,{\rm rad}\), and polarization angle \(1.329\,{\rm rad}\).
The ringdown time is referenced to the H1 peak time
\(t_{\rm peak}^{\rm H1}=1420878141.2190118\), and the ringdown analysis interval is taken to have duration \(T=0.25\,{\rm s}\).

We use the same spectral prior ranges for all agnostic DS analyses considered in this work.
For each DS component, the frequency and damping time are assigned uniform priors,
\begin{equation}
f_k \sim \mathcal{U}(140,280)\,{\rm Hz},
\qquad
\tau_k \sim \mathcal{U}(5\times10^{-4},10^{-2})\,{\rm s}.
\label{eq:spectral_priors}
\end{equation}
For multi-component DS analyses, we impose a prior ordering on the damping times to break the label degeneracy between otherwise exchangeable DS components.

For comparison with the analytically marginalized inference, we also consider direct amplitude--phase sampling.
In the direct-amplitude analyses, the phase is assigned a uniform prior,
\begin{equation}
\phi_k \sim \mathcal{U}(0,2\pi),
\end{equation}
while the amplitude prior is chosen either to be uniform in amplitude,
\begin{equation}
A_k \sim \mathcal{U}(10^{-23},10^{-19}),
\end{equation}
or uniform in log-amplitude,
\begin{equation}
\log_{10} A_k \sim \mathcal{U}(-23,-19).
\end{equation}
In the analytically marginalized analyses, the amplitude--phase coefficients are not sampled explicitly.
Instead, each DS component \(k\) is assigned a Gaussian prior scale \(\sigma_k\), estimated in the pre-merger-informed layer of Sec.~\ref{sec:premerger} and assigned to the agnostic DS components according to the convention described in Sec.~\ref{sec:analytic_marginalization}.
For clarity, we compare the corresponding induced prior densities on the amplitude \(A_k\) in Appendix~\ref{app:amplitude_priors}.

For the theory-side Kerr remnant inference, we use uniform priors on the detector-frame remnant mass and dimensionless remnant spin,
\begin{equation}
M_f \sim \mathcal{U}(40,100)\,M_\odot,
\qquad
\chi_f \sim \mathcal{U}(0.1,0.9).
\label{eq:remnant_priors}
\end{equation}
The Kerr QNM frequencies and damping times entering the theory-side spectral map are computed using the \texttt{qnm} package~\cite{Stein:2019mop}.

All inference runs are performed with \texttt{cpnest}~\cite{veitchJohnveitchCpnestMinor2017}, using \(\texttt{nlive}=2048\) and \(\texttt{maxmcmc}=2048\).
For the agnostic spectral-inference layer, we quantify the Bayesian support for the DS signal model relative to the noise-only model using the log Bayes factor
\begin{equation}
\ln B
=
\ln Z_{\rm signal}
-
\ln Z_{\rm noise},
\label{eq:lnB_signal_noise}
\end{equation}
where \(Z_{\rm signal}\) and \(Z_{\rm noise}\) are the evidences of the DS signal model and the corresponding noise-only model, respectively.

\section{Application and validation}
\label{sec:application_validation}

We now apply \texttt{SPRING} to GW250114 and validate the main steps of the framework.
The goals of this section are twofold.
First, we demonstrate how the pre-merger-informed amplitude scale, the spectral likelihood function, and the theory-side Kerr remnant inference are combined in practice.
Second, we assess whether the resulting spectral and remnant-inference conclusions are stable under changes of the pre-merger input, the amplitude scale, and the ringdown fitting start time.

Throughout this section, we use a 2DS model whose components are associated, under the identification described in Sec.~\ref{sec:spectral_inference}, with the selected QNM content \(\{220,221\}\).

\subsection{Pre-merger-informed amplitude scales}
\label{sec:app_amplitude_scales}

\begin{figure*}
\centering
\includegraphics[
    width=0.95\textwidth
]{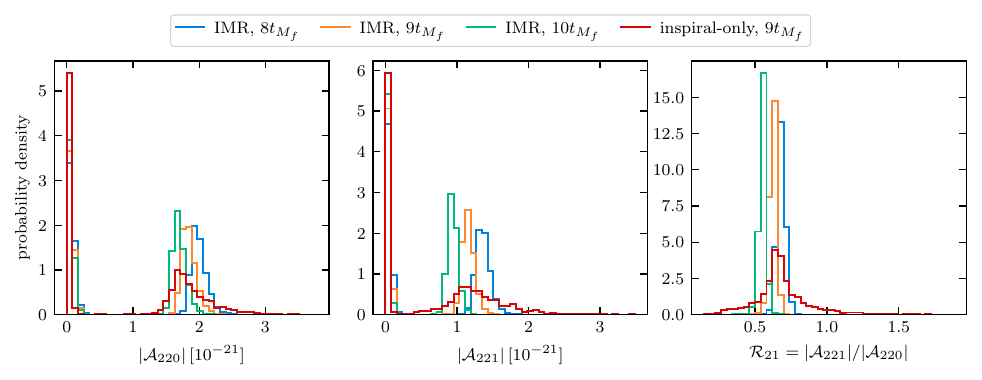}
\caption{
Distributions of the fitted complex-amplitude magnitudes
\(|\mathcal{A}_{220}|\), \(|\mathcal{A}_{221}|\), and the magnitude ratio
\(\mathcal{R}_{21}=|\mathcal{A}_{221}|/|\mathcal{A}_{220}|\)
obtained in the pre-merger-informed layer using either IMR or inspiral-only posterior samples.
All histograms are normalized to probability densities.
}
\label{fig:amp_scales}
\end{figure*}

We first construct the pre-merger-informed amplitude scales used in the spectral-inference layer.
For the main analysis, these scales are constructed from the publicly available GWOSC IMR posterior samples obtained with the \texttt{SEOBNRv5PHM} waveform model
\cite{Pompili:2023tna,
khalil2023theoreticalgroundworksupportingprecessingspin,
vandemeent2023enhancingseobnrv5effectiveonebodywaveform,
Ramos-Buades:2023ehm}.
As a cross-check, we also repeat the amplitude extraction using inspiral-only posterior samples publicly released in Ref.~\cite{collaborationGW250114TestingHawkings2025}.
These samples were obtained with the \texttt{tdinf} package~\cite{Miller_2024,
Miller_2025,
tdinf} from an inspiral-only analysis that uses data only up to \(40t_M\) before merger.
Here \(t_M=GM/c^3\), with \(M\) denoting the binary total mass.

For each posterior sample used in the amplitude-scale construction, we generate the corresponding time-domain waveform modes using \texttt{pyseobnr}, following the procedure described in Sec.~\ref{sec:premerger}.
The waveform peak time \(t_{\rm peak}\) is defined using the mode norm in Eq.~\eqref{eq:tpeak_mode_norm}.
In the present application, \(\mathcal{M}_{\rm peak}\) includes all available modes with \(\ell=2,3,4\), namely
\begin{equation}
\mathcal{M}_{\rm peak}
=
\{(\ell,m): \ell\in\{2,3,4\},\; m=-\ell,\ldots,\ell\}.
\label{eq:Mpeak_application}
\end{equation}
The target waveform for amplitude extraction is the projected \((\ell,m)=(2,2)\) waveform, and the QNM basis is chosen as
\(\mathcal{J}=\{(2,2,0),(2,2,1)\}\).
For each target waveform generated from these posterior samples, we perform the complex least-squares fit described in Sec.~\ref{sec:premerger} and obtain the fitted complex amplitudes \(\mathcal{A}_{220}\) and \(\mathcal{A}_{221}\).
Appendix~\ref{app:qnm_fit_residuals} examines the least-squares residuals of these QNM amplitude fits as a diagnostic of the fit quality.

We define the complex-amplitude magnitude ratio as
\begin{equation}
\mathcal{R}_{21}\equiv
\frac{|\mathcal{A}_{221}|}{|\mathcal{A}_{220}|}.
\label{eq:R21_def}
\end{equation}
Figure~\ref{fig:amp_scales} shows the distributions of
\(|\mathcal{A}_{220}|\), \(|\mathcal{A}_{221}|\), and \(\mathcal{R}_{21}\) for the fitting start time
\(t_0=t_{\rm peak}+9t_{M_f}\), using either IMR or inspiral-only posterior samples as input to the amplitude-scale construction.
The individual amplitude magnitude distributions are not well described by simple Gaussian distributions, whereas the magnitude ratio \(\mathcal{R}_{21}\) is much more concentrated and approximately unimodal.
This suggests that, in this application, the relative excitation strength between the QNM components, as encoded in \(\mathcal{R}_{21}\), is comparatively stable, even though the absolute scale of the fitted amplitudes can be sensitive to the intrinsic binary parameters and extrinsic source parameters sampled from the upstream posterior.

Following Sec.~\ref{sec:premerger}, we define the amplitude scale for each selected QNM component as the median of the fitted complex-amplitude magnitude distribution.
For the main \(t_0=t_{\rm peak}+9t_{M_f}\) analysis, this gives
\begin{equation}
\sigma_{220}
=
1.72\times10^{-21},
\qquad
\sigma_{221}
=
1.05\times10^{-21}.
\label{eq:amp_scales_application}
\end{equation}
These scales are assigned to the two agnostic DS components according to the damping-time ordering described in Sec.~\ref{sec:analytic_marginalization}, namely
\(\sigma_1=\sigma_{220}\) and \(\sigma_2=\sigma_{221}\).

\begin{table}[t]
\centering
\caption{
Pre-merger-informed amplitude information extracted from the QNM fits.
The upper block uses IMR posterior input, while the lower block uses inspiral-only posterior input.
}
\renewcommand{\arraystretch}{1.15}
\setlength{\tabcolsep}{5.5pt}
\begin{tabular}{cccc}
\hline
\(t_0-t_{\rm peak}\) & \(\sigma_{220}\) & \(\sigma_{221}\) & \({\rm median}(\mathcal{R}_{21})\) \\
\hline\hline
\multicolumn{4}{c}{\textit{IMR posterior}} \\
\hline
\(8t_{M_f}\)  & \(1.86\times10^{-21}\) & \(1.24\times10^{-21}\) & 0.68 \\
\(9t_{M_f}\)  & \(1.72\times10^{-21}\) & \(1.05\times10^{-21}\) & 0.63 \\
\(10t_{M_f}\) & \(1.58\times10^{-21}\) & \(8.52\times10^{-22}\) & 0.55 \\
\hline\hline
\multicolumn{4}{c}{\textit{Inspiral-only posterior}} \\
\hline
\(9t_{M_f}\)  & \(1.48\times10^{-21}\) & \(6.41\times10^{-22}\) & 0.66 \\
\hline
\end{tabular}
\label{tab:amp_scales}
\end{table}

Table~\ref{tab:amp_scales} summarizes the amplitude scales extracted at different fitting start times and from different upstream posterior inputs.
For the IMR posterior input, when the fitting start time is shifted from \(8t_{M_f}\) to \(9t_{M_f}\) and \(10t_{M_f}\) relative to \(t_{\rm peak}\), the median of
\(\mathcal{R}_{21}=|\mathcal{A}_{221}|/|\mathcal{A}_{220}|\) decreases.
This trend is consistent with the faster decay of the \(221\) overtone relative to the \(220\) fundamental mode.
The inspiral-only result is also shown as a cross-check of the amplitude-scale construction using an alternative upstream posterior input.

In the subsequent spectral-inference layer, we use these extracted amplitude scales only as soft prior scales, not as deterministic amplitude values.
The relative stability of \(\mathcal{R}_{21}\) motivates the robustness tests below.
In Sec.~\ref{sec:app_robustness}, we repeat the inference under the rescalings
\[
\sigma_k \rightarrow 0.75\,\sigma_k,\qquad
\sigma_k \rightarrow \sigma_k,\qquad
\sigma_k \rightarrow 1.5\,\sigma_k .
\]

\subsection{Agnostic spectral inference}
\label{sec:app_spectral_inference}

We next apply the spectral-inference layer to the GW250114 ringdown data.
For the ringdown data analysis itself, we define a fixed mass-time conversion
\(t_{M_f}^{\rm RD}=0.337\,{\rm ms}\), based on a representative detector-frame remnant mass for GW250114.
In this subsection, we focus on a ringdown fitting start time corresponding to a \(9t_{M_f}^{\rm RD}\) shift relative to the H1 peak time \(t_{\rm peak}^{\rm H1}\).
This fixed ringdown-time conversion differs from the pre-merger-informed amplitude extraction, where \(t_{M_f}\) is evaluated separately for each posterior sample used in the amplitude-scale construction.

We compare four agnostic 2DS analyses.
The first two use direct amplitude--phase sampling with broad amplitude priors, namely a prior uniform in \(A_k\) and a prior uniform in \(\log_{10}A_k\).
The remaining two analyses use analytically marginalized amplitude--phase inference, with amplitude scales obtained from the pre-merger-informed layer: one set of scales is constructed from the IMR posterior input, and the other from the inspiral-only posterior input.
For the \(9t_{M_f}^{\rm RD}\) analysis considered here, we use the IMR and inspiral-only amplitude scales extracted at the corresponding \(9t_{M_f}\) offset and reported in Table~\ref{tab:amp_scales}.
In all cases, the spectral variables are inferred agnostically with the prior ranges specified in Sec.~\ref{sec:inference_setup}; no Kerr or other theory-specific spectral relation is imposed in the DS waveform model.

\begin{figure*}
\centering
\includegraphics[
    width=0.65\textwidth
]{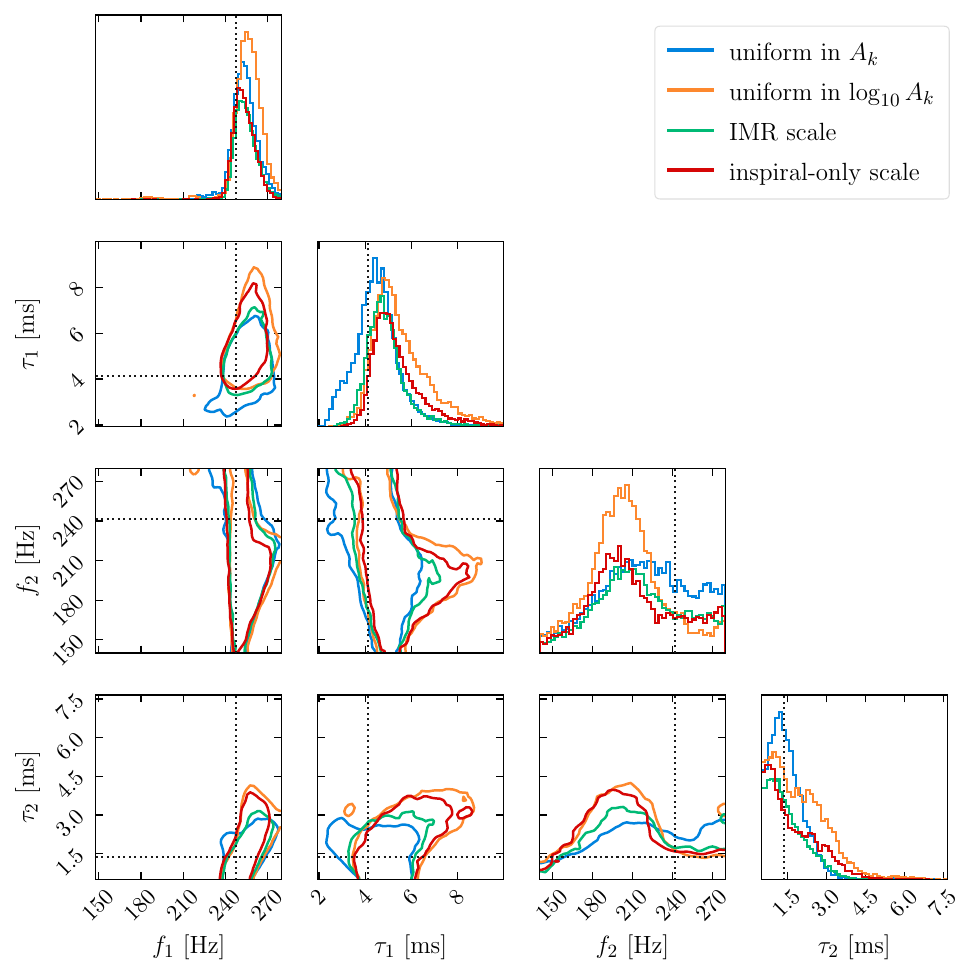}
\caption{
Agnostic 2DS spectral posteriors at a \(9t_{M_f}^{\rm RD}\) ringdown-start offset.
The four curves compare direct amplitude--phase sampling with broad, uniform priors in \(A_k\) and in \(\log_{10}A_k\), and analytically marginalized inference with IMR and inspiral-only amplitude scales.
Black dotted lines indicate the Kerr \(220+221\) QNM spectrum evaluated at the reference IMR remnant parameters.
}
\label{fig:spectral_posterior_9Mf}
\end{figure*}

Figure~\ref{fig:spectral_posterior_9Mf} shows the resulting spectral posterior distributions.
At the level of the marginalized spectral variables shown in the corner plot, the four amplitude treatments give broadly similar posterior support.
Therefore, the improvement introduced by the pre-merger-informed amplitude scales is not most clearly visible as a dramatic shift of the one- and two-dimensional marginalized spectral posteriors.
Instead, the impact of the amplitude-scale information is more clearly reflected in the relative Bayesian support summarized below.

\begin{table}[t]
\centering
\caption{Log Bayes factors for the agnostic 2DS analyses at a \(9t_{M_f}^{\rm RD}\) ringdown-start offset relative to the H1 peak time.}
\renewcommand{\arraystretch}{1.15}
\setlength{\tabcolsep}{18pt}
\begin{tabular}{cc}
\hline
Amplitude treatment & \(\ln B\) \\
\hline
Uniform in \(A_k\) & 247.0 \\
Uniform in \(\log_{10}A_k\) & 251.5 \\
IMR scale & 256.9 \\
inspiral-only scale & 255.9 \\
\hline
\end{tabular}
\label{tab:spectral_lnB}
\end{table}

The corresponding log Bayes factors are summarized in Table~\ref{tab:spectral_lnB}.
We report \(\ln B=\ln Z_{\rm signal}-\ln Z_{\rm noise}\), as defined in Eq.~\eqref{eq:lnB_signal_noise}.
The pre-merger-informed amplitude scales increase the log Bayes factor by \(\Delta\ln B\sim 5\)--\(10\) relative to the analyses with broad, uniform amplitude priors.
This represents a substantial improvement in the Bayesian support for the agnostic 2DS signal model.

This Bayes-factor gain indicates that the amplitude-scale information from the pre-merger-informed layer is informative for the agnostic spectral-inference problem.
The pre-merger-informed scales introduce physically motivated amplitude information and reduce the unnecessary prior volume explored by the amplitude--phase parameters.
At the same time, they do not impose deterministic amplitude values: the induced amplitude prior retains support over all \(A_k\ge0\), providing a soft preference for amplitudes near the pre-merger-informed scale rather than fixing the amplitudes to prescribed values.

\subsection{Kerr remnant inference}
\label{sec:app_kerr_remnant}

\begin{figure}[t]
\centering
\includegraphics[
    width=\columnwidth
]{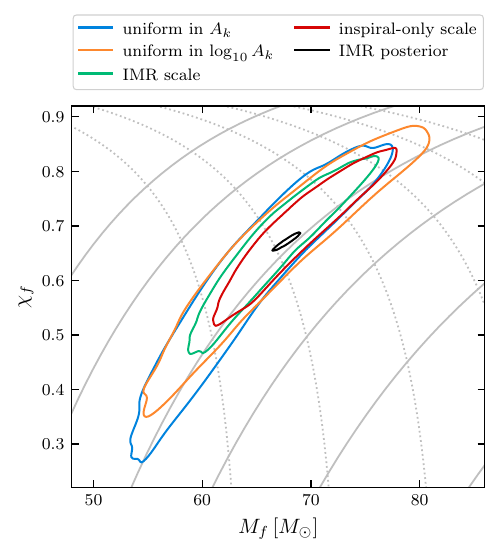}
\caption{
Kerr remnant posteriors inferred from the \(9t_{M_f}^{\rm RD}\) agnostic 2DS spectral-inference outputs.
Colored contours show the \(90\%\) credible regions obtained with different amplitude treatments, together with the reference IMR posterior.
Gray solid and dotted curves indicate constant-frequency and constant-damping-time contours of the Kerr \(220\) QNM mode, respectively.
}
\label{fig:remnant_9Mf_prior_comparison}
\end{figure}

\begin{figure}[t]
\centering
\includegraphics[
    width=\columnwidth
]{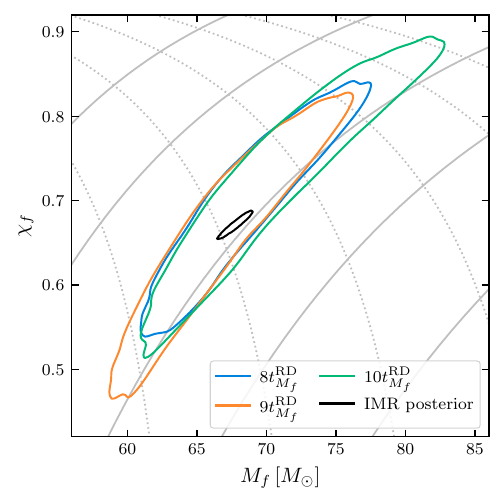}
\caption{
Kerr remnant posteriors inferred from the pre-merger-informed agnostic 2DS spectral likelihood functions at different ringdown-start offsets.
Colored contours show the \(90\%\) credible regions for \(8t_{M_f}^{\rm RD}\), \(9t_{M_f}^{\rm RD}\), and \(10t_{M_f}^{\rm RD}\), using the amplitude scales obtained from the IMR posterior input and listed in Table~\ref{tab:amp_scales}.
The reference IMR posterior is shown in black.
Gray solid and dotted curves indicate constant-frequency and constant-damping-time contours of the Kerr \(220\) QNM mode, respectively.
}
\label{fig:remnant_t0_comparison}
\end{figure}

We now use the spectral likelihood functions constructed in the spectral-inference layer to perform Kerr remnant inference.
For a deterministic Kerr spectral map, the event likelihood is obtained by evaluating the spectral likelihood function on the Kerr-predicted QNM spectrum, as in Eq.~\eqref{eq:event_likelihood_deterministic}.
Here the remnant parameters are \(r=(M_f,\chi_f)\), and the theory-side spectral map is
\(s^{\rm th}=\mathcal{T}_{\rm Kerr}(M_f,\chi_f)\).

In the present \(220+221\) application, the ordered spectral variables inferred by the 2DS model are
\begin{equation}
s=(f_1,\tau_1,f_2,\tau_2),
\end{equation}
where \(k=1\) denotes the longer-lived component and \(k=2\) denotes the shorter-lived component.
Under the component-identification convention used here, the longer-lived component is associated with \(220\), and the shorter-lived component is associated with \(221\).
The Kerr spectral map is therefore evaluated in the corresponding ordered form,
\begin{equation}
s^{\rm th}
=
\mathcal{T}_{\rm Kerr}(M_f,\chi_f)
=
\left(
f_{220}^{\rm th},
\tau_{220}^{\rm th},
f_{221}^{\rm th},
\tau_{221}^{\rm th}
\right),
\label{eq:kerr_spectrum_ordered_app}
\end{equation}
where the dependence of the individual QNM frequencies and damping times on \((M_f,\chi_f)\) is kept implicit on the right-hand side.
The Kerr remnant likelihood is therefore
\begin{equation}
p(d\mid M_f,\chi_f)
=
\mathcal{L}_{\rm spec}
\!\left[s^{\rm th}(M_f,\chi_f)\right].
\label{eq:kerr_remnant_likelihood_app}
\end{equation}

For the analytically marginalized runs, \(\mathcal{L}_{\rm spec}(s)=p(d\mid s)\) is explicitly available from Eq.~\eqref{eq:log_marg_like}.
For the direct amplitude--phase sampling analyses, the sampler instead provides posterior samples over the spectral variables.
For comparison, we reconstruct an approximate spectral likelihood from these posterior samples using
\begin{equation}
p(d\mid s)
\propto
\frac{p(s\mid d)}{p_s(s)},
\label{eq:posterior_to_likelihood_kde}
\end{equation}
where \(p_s(s)\) is the spectral prior used in the agnostic DS inference.
The posterior density \(p(s\mid d)\) is estimated from the spectral posterior samples using kernel-density estimation.

\begin{table}[htbp]
\centering
\caption{
Summary of Kerr remnant posteriors inferred from the \(9t_{M_f}^{\rm RD}\) agnostic 2DS spectral-inference outputs.
Intervals denote the central \(90\%\) credible ranges.
The IMR posterior is included as a reference.
}
\renewcommand{\arraystretch}{1.15}
\setlength{\tabcolsep}{7pt}
\begin{tabular}{ccc}
\hline
Case & \(M_f/M_\odot\) & \(\chi_f\) \\
\hline\hline
\multicolumn{3}{c}{\textit{Kerr remnant inference}} \\
\hline
Uniform in \(A_k\) & \( 64.7^{+8.4}_{-9.6} \) & \( 0.63^{+0.15}_{-0.30} \) \\
Uniform in \(\log_{10}A_k\) & \( 67.7^{+9.4}_{-10.8} \) & \( 0.69^{+0.14}_{-0.26} \) \\
IMR scale & \( 66.6^{+6.5}_{-6.4} \) & \( 0.67^{+0.11}_{-0.17} \) \\
inspiral-only scale & \( 68.8^{+6.5}_{-6.1} \) & \( 0.70^{+0.10}_{-0.14} \) \\
\hline\hline
\multicolumn{3}{c}{\textit{Reference}} \\
\hline
IMR posterior & \( 67.8^{+1.0}_{-1.0} \) & \( 0.67^{+0.01}_{-0.01} \) \\
\hline
\end{tabular}
\label{tab:remnant_9Mf_summary}
\end{table}

Figure~\ref{fig:remnant_9Mf_prior_comparison} compares the Kerr remnant posteriors obtained from the four \(9t_{M_f}^{\rm RD}\) spectral-inference outputs.
Although the marginalized spectral posteriors in Fig.~\ref{fig:spectral_posterior_9Mf} are broadly similar in their one- and two-dimensional projections, the resulting Kerr remnant posteriors show a clearer dependence on the amplitude treatment.
The remnant posteriors inferred from the pre-merger-informed spectral likelihood functions are narrower and more closely aligned with the reference IMR remnant posterior than those obtained from the analyses with broad, uniform amplitude priors.
The results obtained using amplitude scales constructed from the IMR and inspiral-only posterior inputs are mutually similar, indicating that the remnant inference is not strongly driven by the particular posterior input used to construct the amplitude scale.
The corresponding median values and central \(90\%\) credible intervals are listed in Table~\ref{tab:remnant_9Mf_summary}.

\begin{table}[t]
\centering
\caption{
Summary of Kerr remnant posteriors inferred from the pre-merger-informed spectral likelihood functions at different ringdown-start offsets.
The amplitude scales are obtained from the IMR posterior input and listed in Table~\ref{tab:amp_scales}.
Intervals denote the central \(90\%\) credible ranges.
The IMR posterior is included as a reference.
}
\renewcommand{\arraystretch}{1.15}
\setlength{\tabcolsep}{12pt}
\begin{tabular}{ccc}
\hline
Case & \(M_f/M_\odot\) & \(\chi_f\) \\
\hline\hline
\multicolumn{3}{c}{\textit{Kerr remnant inference}} \\
\hline
\(8t_{M_f}^{\rm RD}\) & \( 68.6^{+6.1}_{-6.1} \) & \( 0.71^{+0.10}_{-0.14} \) \\
\(9t_{M_f}^{\rm RD}\) & \( 66.6^{+6.5}_{-6.4} \) & \( 0.67^{+0.11}_{-0.17} \) \\
\(10t_{M_f}^{\rm RD}\) & \( 70.8^{+8.3}_{-7.8} \) & \( 0.74^{+0.11}_{-0.17} \) \\
\hline\hline
\multicolumn{3}{c}{\textit{Reference}} \\
\hline
IMR posterior & \( 67.8^{+1.0}_{-1.0} \) & \( 0.67^{+0.01}_{-0.01} \) \\
\hline
\end{tabular}
\label{tab:remnant_t0_summary}
\end{table}

To check that the agreement with the IMR remnant estimate is not tied to a single choice of ringdown start time, we repeat the Kerr remnant inference for the pre-merger-informed analytically marginalized runs at three start-time offsets,
\(8t_{M_f}^{\rm RD}\), \(9t_{M_f}^{\rm RD}\), and \(10t_{M_f}^{\rm RD}\),
relative to the H1 peak time.
The resulting posteriors are shown in Fig.~\ref{fig:remnant_t0_comparison}, with numerical summaries given in Table~\ref{tab:remnant_t0_summary}.
The three posteriors occupy broadly similar regions of the \((M_f,\chi_f)\) plane and remain consistent with the reference IMR remnant posterior.
The \(10t_{M_f}^{\rm RD}\) result is somewhat broader and extends further toward larger \(M_f\) and \(\chi_f\), consistent with the lower effective ringdown SNR available at later start times.
This indicates that the Kerr remnant inference from the pre-merger-informed spectral likelihood function is reasonably stable under moderate changes of the ringdown fitting start time.

\subsection{Robustness to the amplitude scale}
\label{sec:app_robustness}

\begin{figure}[t]
\centering
\includegraphics[
    width=\columnwidth
]{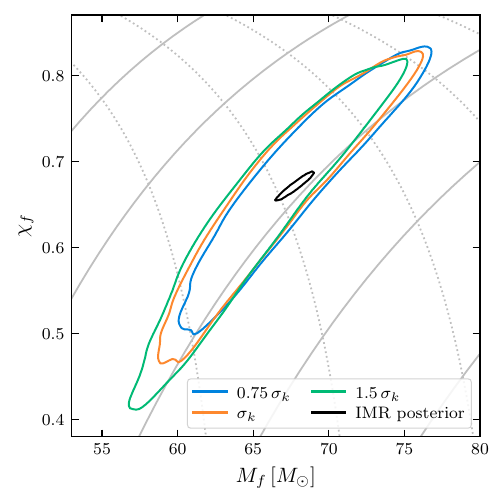}
\caption{
Robustness of the Kerr remnant posterior under global rescalings of the pre-merger-informed amplitude scale constructed from the IMR posterior input.
Colored contours show the \(90\%\) credible regions obtained with amplitude-scale factors \(0.75\sigma_k\), \(\sigma_k\), and \(1.5\sigma_k\) at the \(9t_{M_f}^{\rm RD}\) ringdown-start offset, where \(\sigma_k\) denotes the unrescaled amplitude scale listed in Table~\ref{tab:amp_scales}.
The reference IMR posterior is shown in black.
Gray solid and dotted curves indicate constant-frequency and constant-damping-time contours of the Kerr \(220\) QNM mode, respectively.
}
\label{fig:remnant_amp_scale_robustness}
\end{figure}

We finally assess the robustness of the analysis with respect to the overall scale of the pre-merger-informed amplitude prior.
For this test, we keep the ringdown fitting start time fixed at the \(9t_{M_f}^{\rm RD}\) offset and use the \(9t_{M_f}\) amplitude scales constructed from the IMR posterior input and listed in Table~\ref{tab:amp_scales}.
We then repeat the agnostic spectral inference and the subsequent Kerr remnant inference under the global rescalings
\begin{equation}
\sigma_k \rightarrow 0.75\,\sigma_k,
\qquad
\sigma_k \rightarrow \sigma_k,
\qquad
\sigma_k \rightarrow 1.5\,\sigma_k .
\label{eq:amp_scale_rescaling}
\end{equation}
This rescaling changes the overall amplitude scale while preserving the relative excitation strengths between the two DS components.

Across the three amplitude-scale choices, the one- and two-dimensional marginalized spectral posteriors remain broadly consistent.
The log Bayes factors also differ only mildly, with differences of order \(\Delta\ln B\simeq 0.3\).
Figure~\ref{fig:remnant_amp_scale_robustness} shows the corresponding Kerr remnant posteriors.
The three posteriors remain close to one another and all stay aligned with the reference IMR remnant posterior.
Thus, the moderate global rescalings of the adopted amplitude scale do not qualitatively change either the agnostic spectral inference or the theory-side Kerr remnant inference.
Together with the small variation in the log Bayes factor, this indicates that the improvement in Bayesian support found in Sec.~\ref{sec:app_spectral_inference} is not the result of fine-tuning the overall scale of the amplitude prior. 
This supports the interpretation of the pre-merger-informed amplitude scale as a soft, physically motivated prior scale rather than a finely tuned amplitude constraint.

\section{Discussion}
\label{sec:discussion}

Having introduced the \texttt{SPRING} framework and demonstrated its application to GW250114, we now discuss the interpretation and broader implications of the framework.
Compared with existing ringdown analyses, the most distinctive feature of \texttt{SPRING} is that it explicitly disentangles agnostic ringdown spectral measurement from subsequent theory-side interpretation.
This separation is enabled by the construction of the spectral likelihood function \(\mathcal{L}_{\rm spec}(s)\).
Starting from the marginalized spectral likelihood obtained in the agnostic DS inference layer, \texttt{SPRING} defines \(\mathcal{L}_{\rm spec}(s)\) as a likelihood function on spectral space and then evaluates this function on theory-predicted QNM spectra.

In the present implementation, this connection between the agnostic DS spectral inference and the selected QNM content is made through several modeling choices.
The DS inference is performed in a ringdown interval where the selected QNM description is expected to be meaningful.
The number of DS components is chosen to match the selected QNM content, the DS components are ordered by damping time, and the pre-merger-informed amplitude scales guide the inference toward physically motivated amplitude regions.
Under this identification, the marginalized likelihood \(p(d\mid s)\) obtained from the agnostic DS analysis is interpreted, in the theory-inference layer, as the spectral likelihood function \(\mathcal{L}_{\rm spec}(s)\) associated with the selected QNM content.

This identification should not be interpreted as implying that each measured DS component is exactly the corresponding selected QNM in the data.
Rather, it should be understood as part of the spectral-level modeling assumption of \texttt{SPRING}.
This is analogous in spirit to assumptions such as those made in Kerr-template ringdown analyses
\cite{carulloObservationalBlackHole2019,
collaborationTestsGeneralRelativity2021,
collaborationTestsGeneralRelativity2021a,
siegelRingdownGW190521Hints2023,
maUsingRationalFilters2023,
capanoMultimodeQuasinormalSpectrum2023}, where the post-\(t_0\) data are assumed to be well described by a chosen set of Kerr QNMs.
The difference is that \texttt{SPRING} makes this assumption only after extracting the ringdown spectral information from the data, rather than by imposing the Kerr spectral relations during waveform modeling.
This separation makes the assumption more explicit and provides an intermediate spectral-space object, \(\mathcal{L}_{\rm spec}(s)\), that can be inspected, reused, and evaluated under different theory-side spectral maps.

The GW250114 application provides a useful consistency check of this identification.
After the spectral likelihood function constructed in the spectral-inference layer is evaluated on the Kerr \(220+221\) spectral map, the resulting remnant posterior is closely consistent with the IMR estimate.
This agreement suggests that, for the present event and analysis choices, the agnostic 2DS spectral likelihood contains physically relevant information about the selected \(220+221\) QNM content.
At the same time, this should not be interpreted as requiring the low-dimensional marginalized spectral posteriors themselves to appear directly Kerr-like.
Indeed, in the agnostic spectral-inference results shown in Fig.~\ref{fig:spectral_posterior_9Mf}, some one- and two-dimensional projections, especially those involving the shorter-lived DS component \(k=2\), can appear displaced from the reference Kerr values.
This does not preclude obtaining a physically reasonable remnant posterior, because marginalizing a high-dimensional spectral likelihood onto low-dimensional subspaces can discard correlations and geometric information that are relevant for theory-side inference.

This observation motivates a high-dimensional spectral-manifold viewpoint for BH spectroscopy.
The object relevant for remnant or beyond-Kerr inference is not the low-dimensional marginalized spectral posteriors themselves, but the full spectral likelihood function \(\mathcal{L}_{\rm spec}(s)\).
A theory predicts a spectral manifold embedded in this higher-dimensional spectral space, and inference proceeds by evaluating the likelihood along that manifold.
For the Kerr \(220+221\) case, this manifold is parametrized by \((M_f,\chi_f)\) inside the four-dimensional spectral space \((f_{220},\tau_{220},f_{221},\tau_{221})\).
From this viewpoint, apparent disagreement in a projected spectral posterior does not by itself imply inconsistency with Kerr, nor does apparent agreement in a projection guarantee a faithful test.
A systematic spectral-manifold analysis may therefore be important for future no-hair tests
\cite{meidamTestingNohairTheorem2014,
carulloEmpiricalTestsBlack2018,
isiTestingNohairTheorem2019,
bhagwatRingdownOvertonesBlack2020,
bustilloBlackholeSpectroscopyNohair2021}
and for searches for deviations from GR
\cite{abbottTestsGeneralRelativity2016,
collaborationTestsGeneralRelativity2021,
mezzasomaTheoryagnosticFrameworkInspiral2022,
collaborationTestsGeneralRelativity2021a,
collaborationBlackHoleSpectroscopy2025}, where small beyond-Kerr effects may appear as directions in spectral space that cannot be absorbed by changes in the remnant mass and spin.

The present results also emphasize the value of pre-merger information for ringdown inference.
In our GW250114 application, incorporating pre-merger-informed amplitude scales improves the Bayesian support for the agnostic 2DS signal model.
This is consistent with recent work emphasizing that pre-merger binary dynamics and QNM excitation systematics can play an important role in ringdown inference
\cite{chandraBlackholeRingdownAnalysis2025,
mitmanProbingRingdownPerturbation2025,
wangNonlinearVoiceGW2501142026}.
A natural direction for future work is therefore to develop more systematic ways of propagating pre-merger information into ringdown inference to improve its precision and stability.

Although the present application focuses on a single event, a deterministic Kerr spectral map, and a 2DS model, the formalism itself is more general.
As shown in Sec.~\ref{sec:theory_inference}, the same spectral-likelihood construction applies to multi-event analyses with event-dependent remnant parameters and shared beyond-Kerr parameters.
Beyond the 2DS setting considered here, the framework can be extended to a larger number of spectral components.
The present \(220+221\) configuration represents a useful intermediate case: it contains more spectral information than a single-mode analysis, while avoiding the additional degeneracies and interpretational ambiguities that may arise when more components are introduced.
Applications with larger mode content will require a more careful treatment of component identification, prior structure, and spectral-likelihood geometry.

Finally, the analytic marginalization over the linear amplitude--phase coefficients reduces the effective sampling dimensionality of the data-side spectral inference and improves its computational efficiency.
The resulting spectral likelihood function \(\mathcal{L}_{\rm spec}(s)\) can then be reused under different theory-side spectral maps without repeating the time-domain data analysis, providing a modular route to larger event samples and broader theory spaces.

\section{Conclusion}
\label{sec:conclusion}

In this work, we introduced \texttt{SPRING}, a spectral-level pre-merger-informed framework for black-hole ringdown inference.
The framework is built around three layers.
First, pre-merger information is used to estimate characteristic amplitude scales for a selected set of QNM components.
Second, the ringdown data are analyzed with an agnostic DS model whose linear amplitude--phase coefficients are assigned prior scales from the pre-merger-informed layer, and these coefficients are analytically marginalized to construct the spectral likelihood function \(\mathcal{L}_{\rm spec}(s)\).
Third, remnant or beyond-Kerr parameters are inferred by evaluating this spectral likelihood function on theory-predicted QNM spectra.

We applied \texttt{SPRING} to GW250114 in a 2DS analysis associated with the selected QNM content \(\{220,221\}\).
Compared with analyses using broad, uniform amplitude priors, the pre-merger-informed amplitude scales increased the log Bayes factor for the agnostic 2DS signal model by \(\Delta\ln B\sim 5\)--\(10\).
Using the resulting spectral likelihood function for Kerr remnant inference, we obtained remnant posteriors that are closely consistent with the IMR estimate.
The results obtained using amplitude scales constructed from IMR and inspiral-only posterior inputs are mutually consistent.
We also found that the remnant inference remains reasonably stable under moderate changes of the ringdown fitting start time and under global rescalings of the adopted amplitude scale.

Taken together, these results support the spectral likelihood function as the interface between agnostic spectral inference and theory-side interpretation.
By separating the data-side measurement of the ringdown spectrum from the inference of remnant or beyond-Kerr parameters, \texttt{SPRING} enables different theoretical models to be compared using the same data-side likelihood, through their predicted QNM spectral maps.
Its modular structure also provides a natural route toward multi-event analyses and broader theory spaces.
In this sense, \texttt{SPRING} provides a methodological foundation for spectral-level black-hole spectroscopy in the era of next-generation gravitational-wave observations
\cite{branchesiScienceEinsteinTelescope2023,
evansHorizonStudyCosmic2021,
luoTianQinSpaceborneGravitational2016, 
amaro-seoaneLaserInterferometerSpace2017, 
ruanTaijiProgramGravitationalWave2020}.

\begin{acknowledgments}
This paper employs the following software, listed in alphabetical order: 
\texttt{Astropy}~\cite{collaborationAstropyCommunityPython2013, collaborationAstropyProjectBuilding2018, collaborationAstropyProjectSustaining2022},
\texttt{corner}~\cite{corner}, 
\texttt{cpnest}~\cite{veitchJohnveitchCpnestMinor2017}, 
\texttt{GWpy}~\cite{macleodGWpyPythonPackage2021},
\texttt{H5py}~\cite{collettePythonHDF52013},
\texttt{LALSuite}~\cite{lalsuite, swiglal}, 
\texttt{Matplotlib}~\cite{Hunter:2007}, 
\texttt{NumPy}~\cite{harris2020array},  
\texttt{pandas}~\cite{thepandasdevelopmentteamPandasdevPandasPandas2024}, 
\texttt{PESummary}~\cite{hoyPESummaryCodeAgnostic2021}, 
\texttt{pyseobnr}~\cite{mihaylovPySEOBNRSoftwarePackage2023},
\texttt{qnm}~\cite{Stein:2019mop}, 
\texttt{SciPy}~\cite{2020SciPy-NMeth}, 
and
\texttt{tdinf}~\cite{Miller_2024, Miller_2025, tdinf}.

The authors are grateful to Lorenzo Pompili for the helpful suggestion on \texttt{pyseobnr}.
They also thank Harrison Siegel for the valuable advice on data preprocessing.
This work was supported in part by the National Natural Science Foundation of China under Grant No. 12175108.

\end{acknowledgments}

\section*{Data Availability}

The gravitational-wave strain and \texttt{SEOBNRv5PHM}-based IMR posterior samples used in this work are publicly available from GWOSC.
The inspiral-only posterior samples are publicly available from the LVK data release~\cite{ligoscientificcollaborationvirgocollaborationandkagracollaborationGW250114DiscoveryPaper2025}.
The numerical implementation of \texttt{SPRING} used in this work is being prepared for public release and is not yet publicly available.
Additional materials needed to reproduce the main results are available from the corresponding author upon reasonable request.

\appendix
\renewcommand\theequation{\thesection.\arabic{equation}}
\makeatletter
\@addtoreset{equation}{section}
\makeatother

\section{Dominant waveform-mode diagnostics}
\label{app:dominant_mode}

In Sec.~\ref{sec:premerger}, the target waveform used for QNM amplitude-scale extraction is constructed from the projected \((\ell,m)=(2,2)\) waveform mode.
This choice is motivated by the general expectation from NR studies that, for nearly equal-mass, nonprecessing, quasicircular binaries, the \(\ell=|m|=2\) modes dominate the ringdown emission and the retrograde contribution is subdominant.
Here we provide two windowed diagnostics to assess this expectation for the specific GW250114 application.
The first diagnostic quantifies how well the projected \((2,\pm2)\) waveform reproduces the projected all-mode waveform over the relevant post-peak time windows.
The second diagnostic quantifies the projected strength of the \((2,-2)\) retrograde contribution relative to the \((2,2)\) prograde contribution.

For any mode set \(\mathcal{M}\), we define \(h_{\mathcal{M}}(t)\) by the same projection formula as Eq.~\eqref{eq:htar_projection}, with the summation taken over \((\ell,m)\in\mathcal{M}\).
The peak time \(t_{\rm peak}\) is defined by Eq.~\eqref{eq:tpeak_mode_norm} and is evaluated separately for each posterior sample used in the amplitude-scale construction.
For the diagnostic scan, we define a candidate fitting start time
\begin{equation}
t_0(\Delta t_0)
=
t_{\rm peak}
+
\Delta t_0 ,
\label{eq:app_t0_scan}
\end{equation}
where \(\Delta t_0\) is the post-peak fitting start offset.
For a time window starting at this candidate fitting start time, we define
\begin{equation}
W(\Delta t_0)
=
\left\{
t:
\Delta t_0
\leq
t-t_{\rm peak}
\leq
\Delta t_{\max}
\right\},
\label{eq:app_window}
\end{equation}
with \(\Delta t_{\max}=30\,{\rm ms}\).
Equivalently, this window extends from \(t_0(\Delta t_0)\) to \(t_{\rm peak}+\Delta t_{\max}\).
For two projected complex waveforms \(x(t)\) and \(y(t)\), we use the discrete inner product and norm
\begin{equation}
\langle x,y\rangle_W
=
\sum_{t_a\in W}
x(t_a)y^*(t_a),
\quad
\|x\|_W
=
\left(
\sum_{t_a\in W}
|x(t_a)|^2
\right)^{1/2}.
\label{eq:app_inner_norm}
\end{equation}

\begin{figure}[t]
\centering
\includegraphics[
    width=\columnwidth
]{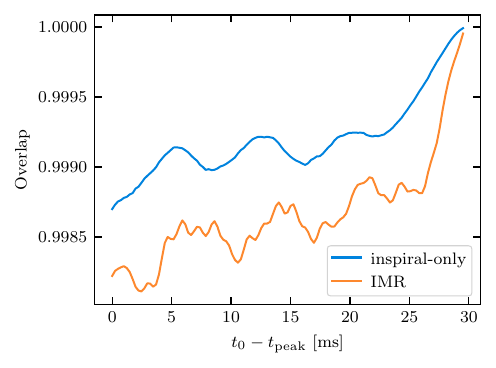}
\caption{
Median windowed overlap between the projected \((2,\pm2)\) waveform and the projected all-mode reference waveform as a function of the fitting start offset \(\Delta t_0=t_0-t_{\rm peak}\).
The window extends from \(t_0(\Delta t_0)\) to \(t_{\rm peak}+30\,{\rm ms}\).
The two curves correspond to waveforms generated using IMR and inspiral-only posterior samples.
}
\label{fig:app_mode_overlap}
\end{figure}

We first define a projected mode-overlap diagnostic.
For the \((2,\pm2)\) comparison mode set \(\mathcal{M}_{2\pm2}\) and an all-mode reference set \(\mathcal{M}_{\rm all}\), the windowed overlap is
\begin{equation}
\mathcal{O}(\Delta t_0)
=
\frac{
\left|
\left\langle
h_{\mathcal{M}_{\rm all}},
h_{\mathcal{M}_{2\pm2}}
\right\rangle_{W(\Delta t_0)}
\right|
}{
\left\|
h_{\mathcal{M}_{\rm all}}
\right\|_{W(\Delta t_0)}
\left\|
h_{\mathcal{M}_{2\pm2}}
\right\|_{W(\Delta t_0)}
}.
\label{eq:app_mode_overlap}
\end{equation}
In the diagnostic used here, we take
\begin{equation}
\mathcal{M}_{2\pm2}
=
\{(2,2),(2,-2)\},
\qquad
\mathcal{M}_{\rm all}
=
\mathcal{M}_{\rm peak}.
\label{eq:app_mode_sets_overlap}
\end{equation}
Here \(\mathcal{M}_{\rm all}\) denotes the mode set used to construct the projected all-mode reference waveform.
In the present implementation, it is taken to be the same as the peak-time mode set \(\mathcal{M}_{\rm peak}\), namely the available modes with \(\ell=2,3,4\) specified in Eq.~\eqref{eq:Mpeak_application}.
Thus, \(\mathcal{O}(\Delta t_0)\) measures how well the projected \((2,\pm2)\) contribution reproduces the projected all-mode reference waveform in the window from \(t_{\rm peak}+\Delta t_0\) to \(t_{\rm peak}+30\,{\rm ms}\).

We also define a projected retrograde-to-prograde strength ratio.
For the reference prograde mode \((\ell,m)=(2,2)\), this ratio is
\begin{equation}
\mathcal{R}_{22}^{\rm retro}(\Delta t_0)
=
\frac{
\left\|
h_{2,-2}^{\rm proj}
\right\|_{W(\Delta t_0)}
}{
\left\|
h_{2,2}^{\rm proj}
\right\|_{W(\Delta t_0)}
},
\label{eq:app_retro_ratio_22}
\end{equation}
where \(h_{2,\pm2}^{\rm proj}\) denotes the individually projected \((2,\pm2)\) waveform mode, constructed using the same projection prescription as Eq.~\eqref{eq:htar_projection}.
This quantity measures the windowed projected strength of the retrograde \((2,-2)\) contribution relative to the prograde \((2,2)\) contribution.

Figure~\ref{fig:app_mode_overlap} shows the median windowed overlap for waveforms generated using either IMR or inspiral-only posterior samples.
For both posterior inputs, the projected \((2,\pm2)\) contribution has an overlap very close to unity with the projected all-mode reference waveform throughout the post-peak range considered.
This indicates that the \((2,\pm2)\) modes dominate the projected waveform over the windows relevant for the amplitude-scale construction.

\begin{figure}[t]
\centering
\includegraphics[
    width=\columnwidth
]{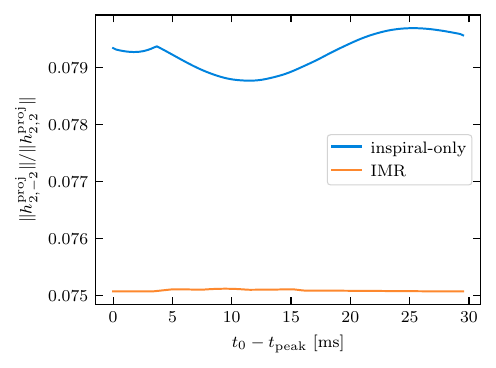}
\caption{
Median windowed projected retrograde-to-prograde strength ratio
\(\mathcal{R}_{22}^{\rm retro}
=
\|h_{2,-2}^{\rm proj}\|_W/
\|h_{2,2}^{\rm proj}\|_W\)
as a function of the fitting start offset \(\Delta t_0=t_0-t_{\rm peak}\).
The window extends from \(t_0(\Delta t_0)\) to \(t_{\rm peak}+30\,{\rm ms}\).
The two curves correspond to waveforms generated using IMR and inspiral-only posterior samples.
}
\label{fig:app_retro_ratio}
\end{figure}

Figure~\ref{fig:app_retro_ratio} shows the median projected retrograde-to-prograde strength ratio.
For both posterior inputs, the projected \((2,-2)\) retrograde contribution remains below \(10\%\) of the projected \((2,2)\) prograde contribution.
This supports the approximation used in Sec.~\ref{sec:premerger}, where the target waveform for QNM amplitude-scale extraction is taken to be the projected \((2,2)\) waveform.

\section{Comparison of induced amplitude priors}
\label{app:amplitude_priors}

\begin{figure*}[t]
\centering
\includegraphics[
    width=0.95\textwidth
]{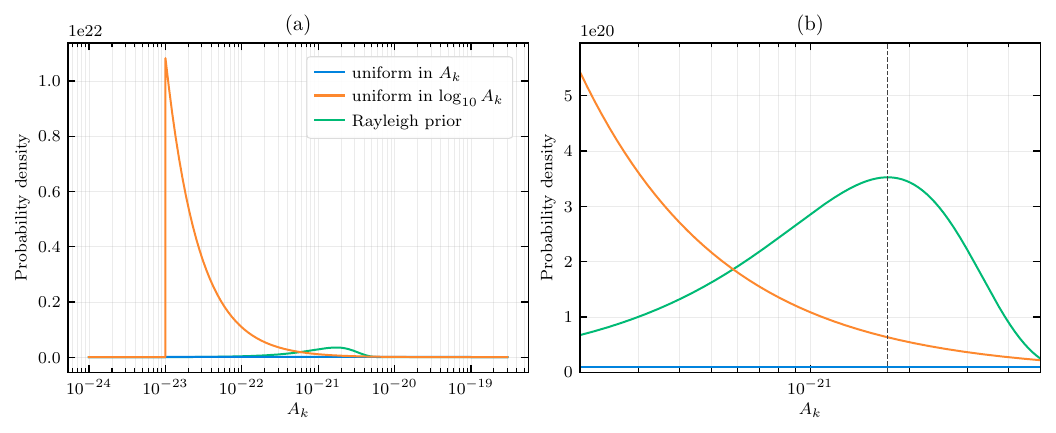}
\caption{
Comparison of induced marginal prior densities on the amplitude \(A_k\), after marginalizing over the phase \(\phi_k\).
The curves show the priors uniform in \(A_k\), uniform in \(\log_{10}A_k\), and the Rayleigh prior induced by a zero-mean Gaussian prior on the linear coefficients \((p_k,q_k)\).
For the two direct-amplitude priors, the amplitude interval is taken to be \(A_k\in[10^{-23},10^{-19}]\), while the Rayleigh prior is shown for \(\sigma_k=1.72\times10^{-21}\).
Panel (b) shows a zoomed-in view of the amplitude range relevant for the pre-merger-informed scale; the black dashed line marks \(A_k=\sigma_k\).
}
\label{fig:app_amplitude_priors}
\end{figure*}

In Sec.~\ref{sec:inference_setup}, we specify three amplitude-prior choices used in the agnostic DS analyses.
The direct amplitude--phase sampling analyses use either a prior uniform in \(A_k\) or a prior uniform in \(\log_{10}A_k\), while the analytically marginalized analyses use a zero-mean Gaussian prior on the linear coefficients \((p_k,q_k)\).
Here we compare the corresponding induced prior densities on the physical amplitude \(A_k\), after marginalizing over the phase \(\phi_k\).

For the direct amplitude prior uniform in \(A_k\), the marginal density is constant within the adopted amplitude range,
\begin{equation}
\pi(A_k)
\propto
1 .
\label{eq:app_prior_uniform_A}
\end{equation}
For the prior uniform in \(\log_{10}A_k\), the induced marginal density on \(A_k\) is
\begin{equation}
\pi(A_k)
\propto
\frac{1}{A_k},
\label{eq:app_prior_uniform_logA}
\end{equation}
within the same amplitude range.
By contrast, the Gaussian prior on the linear coefficients,
\begin{equation}
p_k,q_k \sim \mathcal{N}(0,\sigma_k^2),
\end{equation}
induces a Rayleigh distribution on the amplitude \(A_k=(p_k^2+q_k^2)^{1/2}\),
\begin{equation}
\pi(A_k)
=
\frac{A_k}{\sigma_k^2}
\exp\left[
-\frac{A_k^2}{2\sigma_k^2}
\right],
\qquad
A_k\ge0 .
\label{eq:app_prior_rayleigh}
\end{equation}
The corresponding joint prior in \((A_k,\phi_k)\) is given in Eq.~\eqref{eq:prior_A_phi}.

Figure~\ref{fig:app_amplitude_priors} compares these induced marginal amplitude priors.
The uniform-in-\(A_k\) prior assigns equal density per unit amplitude, while the uniform-in-\(\log_{10}A_k\) prior places substantially more prior density toward small amplitudes.
The Rayleigh prior induced by the Gaussian prior on \((p_k,q_k)\) is regular at \(A_k=0\) and peaks at \(A_k=\sigma_k\), thereby introducing the pre-merger-informed amplitude scale as a soft preference rather than a fixed amplitude value.
This illustrates how the analytically marginalized analysis incorporates amplitude-scale information without imposing a deterministic amplitude constraint.

\section{Residuals of the QNM amplitude fits}
\label{app:qnm_fit_residuals}

\begin{figure}[t]
\centering
\includegraphics[
    width=\columnwidth
]{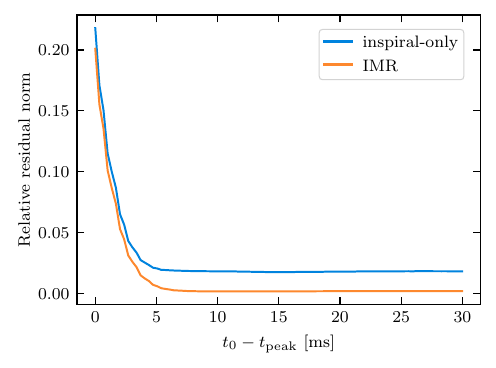}
\caption{
Median relative residual norm of the complex least-squares QNM amplitude fits as a function of the fitting start offset \(\Delta t_0=t_0-t_{\rm peak}\).
The target waveform is the projected \((2,2)\) waveform, and the fitting basis is \(\mathcal{J}=\{(2,2,0),(2,2,1)\}\).
The two curves correspond to waveforms generated using IMR and inspiral-only posterior samples.
}
\label{fig:app_qnm_fit_residuals}
\end{figure}

In the pre-merger-informed layer, the amplitude scales are extracted by fitting a selected QNM basis to a projected target waveform.
Here we examine the residuals of these complex least-squares fits as a diagnostic of the fit quality.

For this diagnostic, we use the same target waveform-mode set \(\mathcal{M}_{\rm tar}\) as in Sec.~\ref{sec:premerger}.
The projected target waveform is
\begin{equation}
h_{\rm tar}(t)
=
h_{\mathcal{M}_{\rm tar}}(t),
\end{equation}
where \(h_{\mathcal{M}_{\rm tar}}(t)\) is constructed using the projection prescription in Eq.~\eqref{eq:htar_projection}.
In the application considered here, we take
\begin{equation}
\mathcal{M}_{\rm tar}
=
\{(2,2)\}.
\end{equation}

We use the same diagnostic scan of candidate fitting start times \(t_0(\Delta t_0)\) defined in Eq.~\eqref{eq:app_t0_scan}, with \(\Delta t_0\in[0,30]\,{\rm ms}\).
For each \(t_0\), we fit the target waveform over the post-\(t_0\) interval using the selected QNM basis.
For a QNM component \(j\in\mathcal{J}\), the basis function is
\begin{equation}
b_j(t-t_0)
=
\exp\left[-i\omega_j(t-t_0)\right],
\qquad
t\geq t_0,
\end{equation}
where the complex QNM frequency \(\omega_j\) is evaluated at the remnant mass and spin associated with the same posterior sample.
In the present diagnostic, the QNM content is
\begin{equation}
\mathcal{J}
=
\{(2,2,0),(2,2,1)\}.
\end{equation}

At discrete times \(t_a\geq t_0\), the complex least-squares problem is
\begin{equation}
h_{\rm tar}(t_a)
\simeq
\sum_{j\in\mathcal{J}}
\mathcal{A}_j(t_0)\,
b_j(t_a-t_0).
\end{equation}
Equivalently, in matrix form,
\begin{equation}
\mathbf{h}_{\rm tar}
\simeq
\mathbf{B}(t_0)\boldsymbol{\mathcal{A}}(t_0),
\end{equation}
with
\begin{equation}
B_{aj}(t_0)
=
b_j(t_a-t_0).
\end{equation}
The fitted complex amplitudes are obtained from
\begin{equation}
\widehat{\boldsymbol{\mathcal{A}}}(t_0)
=
\arg\min_{\boldsymbol{\mathcal{A}}}
\left\|
\mathbf{h}_{\rm tar}
-
\mathbf{B}(t_0)\boldsymbol{\mathcal{A}}
\right\|_2^2 .
\end{equation}

We quantify the fit quality using the relative residual norm
\begin{equation}
\mathcal{R}_{\rm tar}(t_0)
=
\frac{
\left\|
\mathbf{h}_{\rm tar}
-
\mathbf{B}(t_0)
\widehat{\boldsymbol{\mathcal{A}}}(t_0)
\right\|_2
}{
\left\|
\mathbf{h}_{\rm tar}
\right\|_2
}.
\label{eq:app_qnm_fit_residual}
\end{equation}
This quantity measures the fractional mismatch between the projected target waveform and its best-fit \(220+221\) QNM representation over the fitting interval.

Figure~\ref{fig:app_qnm_fit_residuals} shows the median relative residual norm as a function of the fitting start offset \(\Delta t_0=t_0-t_{\rm peak}\), using waveforms generated from either IMR or inspiral-only posterior samples.
For both posterior inputs, the residual decreases rapidly after the peak and becomes small once the fitting start time is shifted a few milliseconds into the post-peak regime.
This behavior supports the use of the \(220+221\) QNM basis for extracting characteristic amplitude scales over the post-peak fitting windows used in the main analysis.


\bibliographystyle{apsrev4-2}
\bibliography{ref}

\end{document}